# Millimeter Wave Communications for Future Mobile Networks

Ming Xiao, *Senior Member, IEEE,* Shahid Mumtaz, *Senior Member, IEEE,*
Yongming Huang, *Senior Member, IEEE,* Linglong Dai, *Senior Member, IEEE,* Yonghui Li, *Senior Member, IEEE,*
Michail Matthaiou, *Senior Member, IEEE,* George K. Karagiannidis, *Fellow, IEEE,* Emil Björnson, *Member, IEEE,*
Kai Yang, *Senior Member, IEEE,* Chih-Lin I., *Senior Member, IEEE,* Amitava Ghosh, *Fellow, IEEE*

***Abstract*—Millimeter wave (mmWave) communications have recently attracted large research interest, since the huge available bandwidth can potentially lead to rates of multiple Gbps (gigabit per second) per user. Though mmWave can be readily used in stationary scenarios such as indoor hotspots or backhaul, it is challenging to use mmWave in mobile networks, where the transmitting/receiving nodes may be moving, channels may have a complicated structure, and the coordination among multiple nodes is difficult. To fully exploit the high potential rates of mmWave in mobile networks, lots of technical problems must be addressed. This paper presents a comprehensive survey of mmWave communications for future mobile networks (5G and beyond). We first summarize the recent channel measurement campaigns and modeling results. Then, we discuss in detail recent progresses in multiple input multiple output (MIMO) transceiver design for mmWave communications. After that, we provide an overview of the solution for multiple access and backhauling, followed by analysis of coverage and connectivity. Finally, the progresses in the standardization and deployment of mmWave for mobile networks are discussed.**

***Index Terms*—Millimeter wave communications, mobile networks, channel model, MIMO beamforming, multiple access, standardization.**

## I. Introduction and Background

With the fast development of electronic devices and computer science, various emerging applications (e.g., virtual reality, augmented reality, big data analytics, artificial intelligence, three-dimensional (3D) media, ultra-high definition transmission video, etc.) have entered our society and created a significant growth in the data volume of wireless networks. Meanwhile, mobile networks have become indispensable to our society as a key service for personal-computing devices. One of the main characteristics of future mobile networks (5G and beyond) is the unprecedented traffic volumes, with huge area spectral efficiency (hundreds of bit/s/Hz/km$^2$) and the very high throughput per device (multiple Gbps). For instance, it is predicted that the world monthly traffic of smartphones will be about 50 petabytes in 2021 [1], which is about 12 times of the traffic in 2016.

In order to meet these requirements, the research and deployment for the future mobile networks [2]–[4] have already been launched. Since 2013, the national-level 5G research organizations and projects (including European Union (EU) 5GPPP/METIS, China IMT-2020 (5G) Promotion Group, Korea 5G Forum, and Japan ARIB) have been set up one after the other to achieve the 2020 technical targets. In 2015, ITU-R officially named 5G systems as IMT-2020, and released recommendation on its framework and overall objectives. Currently, Phase-1 of 5G is being standardized in 3GPP (http://www.3gpp.org/news-events/3gpp-news). Fig. 1 illustrates potential usage scenarios and capabilities of IMT-2020 [4]. Note that 5G is envisaged not only to expand and support diverse usage scenarios and applications that will continue beyond the current networks, but also to support a broad variety of new application scenarios, including:

1) Enhanced Mobile BroadBand (eMBB);
2) Massive machine type communications (mMTC);
3) Ultra Reliable Low Latency Communication (URLLC).

It is expected that IMT-2020 can provide the following eight key performance indicators (KPIs) [4]: greater than 10 Gbit/s peak data rate, 100 Mbit/s user-experienced data rate, 3x spectrum efficiency, greater than 100 Mbps cell edge rates, 10 Mbit/s/km$^2$ area traffic capacity, 100x network energy efficiency, 1 ms over-the-air latency, support for 500 km/h mobility, and $10^6$ km$^2$ connection density [2]. The multiplicative improvements are measured with respect to IMT-Advanced. Recently, EU has also launched Beyond 5G research within the H2020 framework (ICT 2017-09 Call), where a key technology foundation is to use the millimeter wave (mmWave) bands from 30 GHz to 300 GHz, and also THz frequency bands.

To achieve the magnificent objectives and visions listed above, several key enabling technologies have been identified, such as mmWave communications, massive multiple-input and multiple-output (MIMO), small cell deployment, full-duplex relaying, D2D communications, interference management techniques, dynamic TDD with self-backhauling and novel access technologies. Many of these technologies have complementary benefits and need to be combined to achieve all

M. Xiao is with the department of information science and engineering, School of Electrical Engineering, Royal Institute of Technology, Sweden. Email: mingx@kth.se; Shahid Mumtaz, is with Campus Universitario de Santiago, Portugal, Email: smumtaz@av.it.pt; Yongming Huang, is with Southeast University, China, Email: huangym@seu.edu.cn; Linglong Dai is with Tsinghua University, China, Email: daill@tsinghua.edu.cn; Yonghui Li, University of Sydney, Australia, Email: yonghui.li@sydney.edu.au Michail Matthaiou is with Queen's University Belfast, UK, Email: m.matthaiou@qub.ac.uk; George K. Karagiannidis is with Aristotle University of Thessaloniki, Greece, Email: geokarag@auth.gr; Emil Björnson is with Linköping University, Sweden, Email: emil.bjornson@liu.se; Kai Yang is with TongJi University, China, kaiyang@tongji.edu.cn; Chih-Lin I., China Mobile, China, Email: icl@chinamobile.com; Amitava Ghosh, Nokia, USA, Email: amitava.ghosh@nokia.com

Ming Xiao would like to thank Mr. Zhengquan Zhang for valuable discussion and drawing some of the tables.

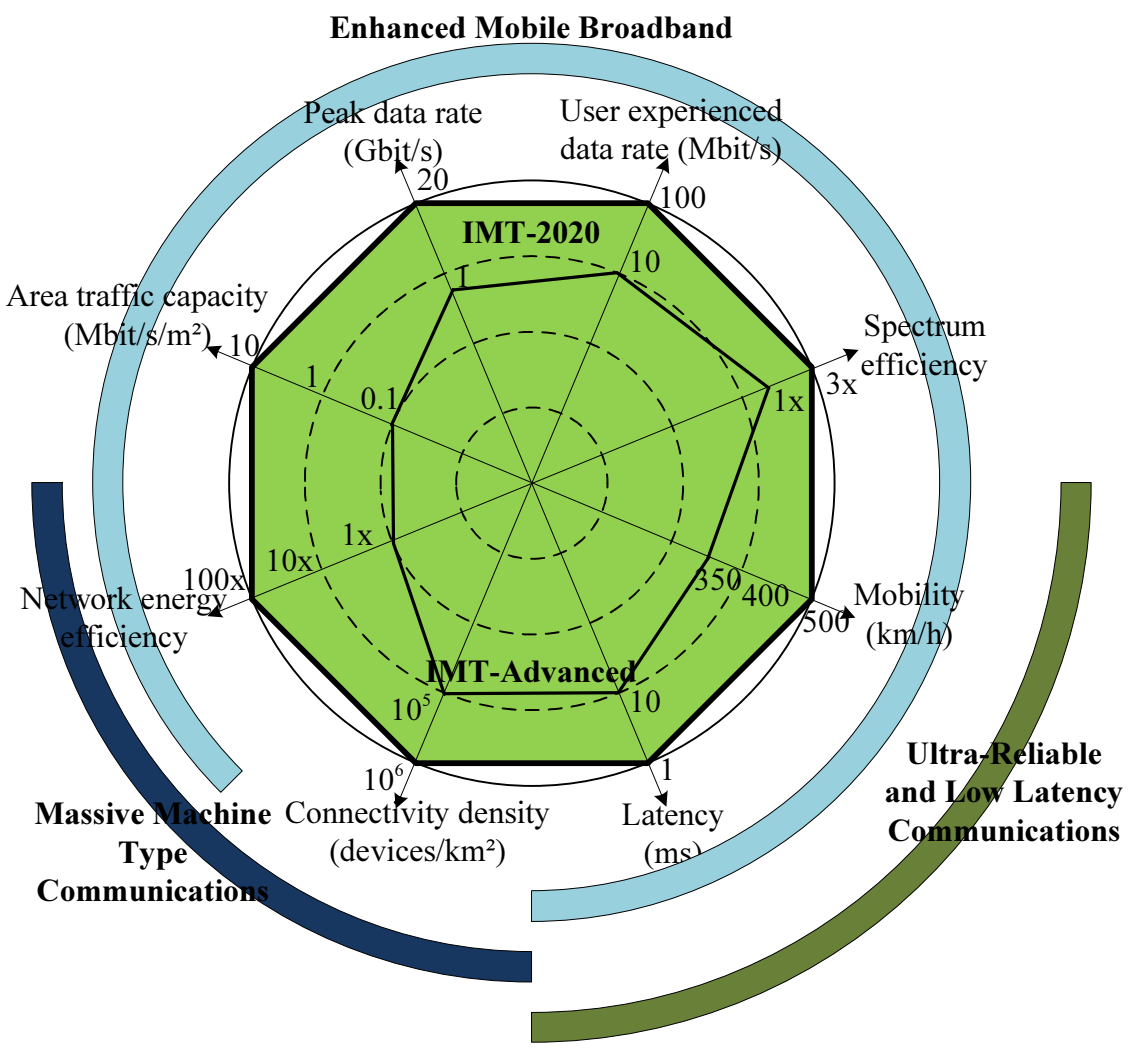

Fig. 1. 5G usage scenarios and key capabilities of IMT-2020, as compared to IMT-Advanced [2].

the key capabilities of 5G. For example, mmWave communications [4], [5], [7], [9], [12], [13] is widely considered the most important technologies to achieve 10 Gbit/s peak data rates. This is because there are a large amount of bandwidth available in the mmWave bands, and expanding the bandwidth is an efficient approach to enhance system capacity. In particular, the channel capacity of an additive white Gaussian noise channel operating over $B$ Hz is

$$C = B \log_2 \left( 1 + \frac{P}{N_0 B} \right), \tag{1}$$

where $P$ is the signal power and $N_0$ is the noise power spectral density [15]. Hence, the capacity increases linearly with the bandwidth $B$, if we also let $P$ grow proportionally to $B$. Since $P$ is limited by regulations in practice, mmWave communication is particularly well-suited for scenarios with good channel conditions, such as short-range small cell access and line-of-sight backhauling in mobile networks [16]–[19]. Self-bcakhaul, where the same wireless spectrum is shared between access and backhaul [20], [23], can provide flexible and cost-efficient solutions to overcome the difficulty of deploying dedicated backhaul, especially in an ultra dense network (UDN) [24]–[26]. It refers to a set of solutions where small base stations (BSs) without dedicated backhaul connect to other BSs that have dedicated backhaul link, by utilizing a similar radio access technology as the one used by the user equipments (UEs) to access the network.

The study of mmWave can trace back to more than 100 years ago. For instance, the experiments at wavelengths as short as 5 and 6 mm were performed by Bose and Lebedew in the 1890s [27]. As for its application in radio communications, the mmWave mobile communications were originally invented in the 1990s and early 2000s, including system design and channel measurements [28]–[30]. Due to the fact that the spectral resource below 6 GHz is becoming scaring while abundant bandwidth is availabe at mmWave band, recently large efforts have been devoted to mmWave communications research and mmWave wireless local area networks (WLAN); for example, IEEE 802.11ad technology operating at 60 GHz, is already available. The more challenging development of mmWave mobile communications is ongoing and is what this paper focuses on. Samsung first achieved 1 Gb/s data transmission at 28 GHz in May 2013. Google also put substantial research efforts into mmWave communications. Verizon has submitted applications to the Federal Communications Commission (FCC) to obtain special temporary authorization (STA) to test mmWave communications technology at 28 GHz. T-Mobile is also expected to obtain STA at 28 GHz and 39 GHz. Nokia in collaboration with National Instruments achieved a peak rate of 15 Gbps using their proof-of-concept system at 73GHz band in April 2015. To further promote the development of mmWave mobile communications, Millimetre-wave evolution for backhaul and access (MiWEBA) (http://www.miweba.eu/, a joint EU-Japan project), Beyond 2020 heterogeneous wireless networks with mmWave small cell access and backhauling (MiWaves) (http://www.miwaves.eu/), and mmWave based mobile radio access network for fifth generation integrated communications (mmMAGIC) (https://5g-mmmagic.eu/) projects have been initiated by EU. In Japan, to prepare for 2020 Tokyo Olympic, DOCOMO and Ericsson tested at 15GHz to reach the rates of 4.5Gbps in outdoor environments and at 70GHz to reach the rates of 2Gbps in indoor environments (https://www.nttdocomo.co.jp/english/info/media_center/pr/2015/030203.html). In China, the minister of science and technology (MOST) has supported a few projects (through project 863) in mmWave in mobile networks and the RF chips of 60GHz and 42-48GHz have been produced (http://www.most.gov.cn/kjbgz/201609/t20160923_127867.htm). Huawei and China mobile also demonstrated the Ka-band (26.5–40GHz) mobile access with 20Gbps rates in the Mobile World Congress 2017.

The mmWave communication has a rich history. Recently the main research interest on mmWave is shifting from local area networks to mobile networks. The main focus of this paper is on development of mmWave mobile communications. We acknowledge the initial contribution in this field in the 1990s and 2000s including channel measurement and system design, but we particularly concentrate on its new research contributions for future mmWave mobile networks. Note that though there have been many research progresses in the area, the existing literature mainly focuses on addressing specific technical challenges. A comprehensive survey, which organically combines these numerous but disjointed works and provides the summary of latest research progress, is basically missing.

Note that references [7], [12]–[14], [16] gave excellent overviews of some aspects of mmWave mobile networks from different aspects. Reference [7] is a pioneering overview paper on the mmWave for mobile networks. Yet, as published in 2013, [7] did not contain the technology development of very recent years. Moreover, some topics, such as, mmWave MIMO, multiple-access, backhual and standardization are not included in [7]. Reference [12] is also an overview paper on mmWave 5G paper published in early 2014, which did not include recent progress of the topic. Moreover, topics of mmWave MIMO and standardization are not included in [12]. Reference [16] gives a good brief overview of mmWave for 5G. However, the introduction in [16] does not include technical details as a magazine paper. [14] is a recent overview paper on signal processing for mmWave MIMO. Yet, reference [14] is focused in mmWave MIMO and other topics e.g., recent developments of channel measure/modeling, multiple access, standardization, field test are missing. Reference [13], published in 2014, discussed the mmWave mobile in channel modeling, modulation, network architecture. Yet, [13] did not discuss the topics of mmWave MIMO, standardization, multiple access. Different from these papers, we provide such a systematic survey on the main progress and technical content of mmWave communications for mobile networks, and further discuss some related research issues and challenges. Especially we include the recent technical development of related topics, including the accepted papers of this IEEE JSAC special issue.

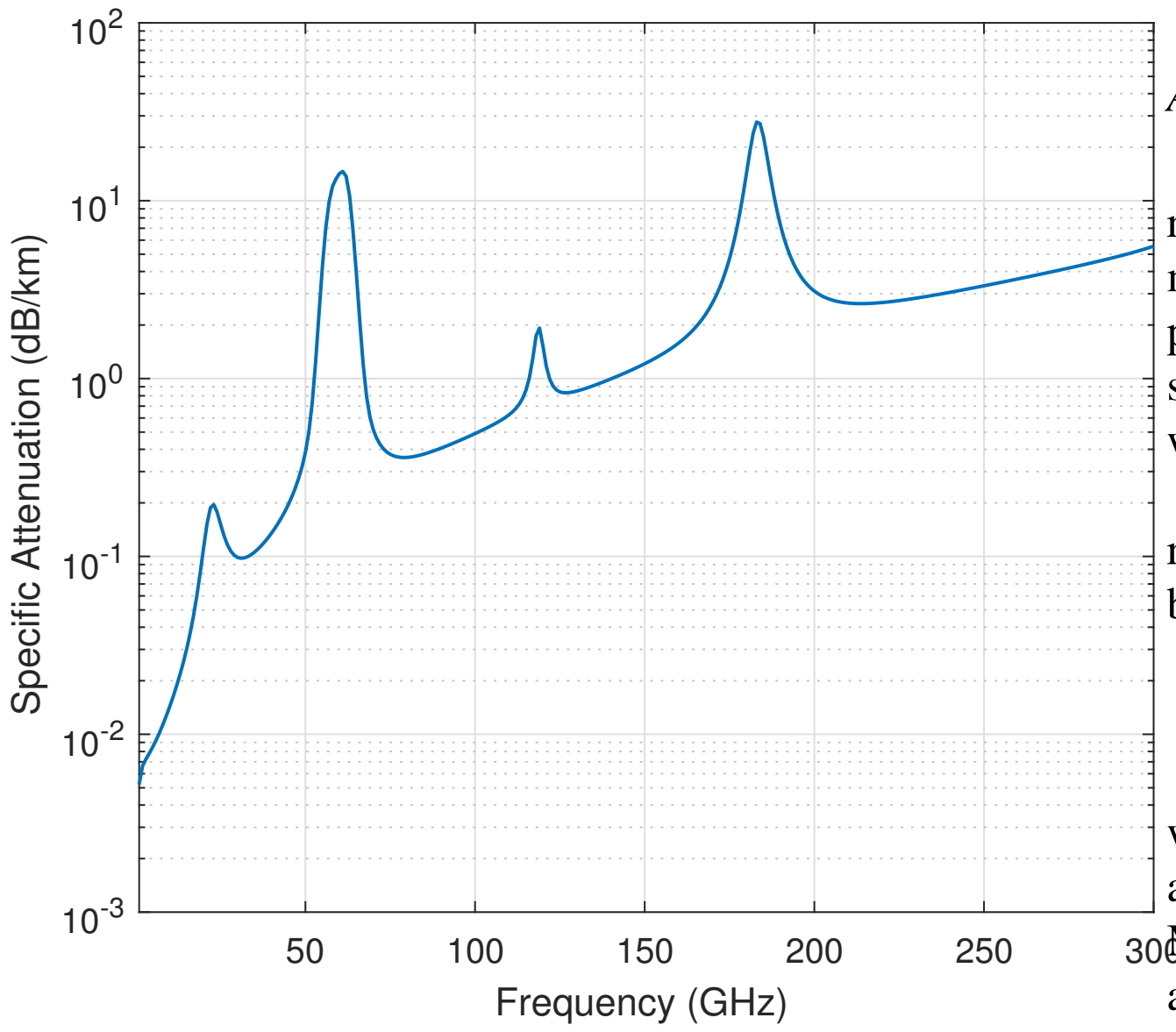

Fig. 2. Atmospheric and molecular absorption in different frequency bands [35].

The rest of the paper is organized as follows. We first give an overview of the key challenges and technical potentials of mmWave communications in future mobile networks in Section II. Then, we discuss mmWave propagation characteristics and channel modeling in Section III, and then present MIMO design for mmWave communications in Section IV. Then, we discuss the multiple access technologies, in-band backhauling, and the coverage performance in Section V. We present the progress of the standardization and deployment of mmWave mobile networks in Section VI. The conclusions are finally given in Section VII.

## II. Key Challenges and Technical Potentials

Next, we will describe some important challenges and potential gains from using mmWave communications in mobile networks.

### A. Main Technical Challenges

Despite the theoretical potentials for extremely high data rates, there are several key technical challenges for using mmWave in mobile networks, including severe pathloss, high penetration loss, high power consumption, blockage due to shadowing, hardware impairment, etc. In what follows, we will give a brief introduction to these topics.

**Pathloss**: In free-space transmission, the power of the received signal (outside the Kirchhoff area) can be determined by *Friis transmission formula* [7]:

$$P_{\mathrm{r}}(d) = P_{\mathrm{t}} G_{\mathrm{t}} G_{\mathrm{r}} \left( \frac{\lambda}{4\pi} \right)^2 d^{-n}, \tag{2}$$

where $P_{\mathrm{t}}$ is the transmit power and $G_{\mathrm{t}}$ and $G_{\mathrm{r}}$ are the antenna gains of the transmitter and receiver, respectively. Moreover, $\lambda$ is wavelength, $d$ is the transmission distance, and the pathloss exponent $n$ equals to 2 in free space. The formula in (2) can be also used to, approximately, describe the power of the received signal in non-free-space propagation as well, by making channel measurements and then finding a suitable value of $n$ that approximately describes the pathloss measurements. The value of $n$ is usually in the range from 2 to 6. There are also refined models, e.g., for cellular networks and $n$ can in some scenarios be smaller than 2 [31].

The wavelength of mmWave signals is much shorter than conventional microwave communication signals, operating at carrier frequency below 6 GHz. Hence, the pathloss of mmWave signals is much higher than that of microwave signals, if all other conditions including the antenna gains are the same. Although the pathloss of mmWave is generally quite high, it is feasible to communicate over the distances that are common in urban mobile networks, such as a few hundreds of meters [7] or even a few kilometers [32]. By using directive antennas, it has been demonstrated that 10 km communication ranges are possible under clean air conditions [32]. If the air is not clean, the rain attenuation and atmospheric/molecular absorption increase the pathloss and limit the communication range [33], [34]. The impact of these factors varies with the carrier frequency; for instance, the atmospheric and molecular absorption are shown in Fig. 2 and the rain attenuation is shown in Fig. 3.

**Penetration loss**: The pathloss discussion assumes line-of-sight (LoS) communications, but the high penetration loss is compounded in non-line-of-sight (NLoS) scenarios. In indoor environments, although the penetration losses for clear glass and dry walls are relatively low for 28 GHz signals (comparable to microwave bands), the penetration losses for brick and tinted glass are high for 28 GHz signals (about 28 dB and 40 dB), which is much higher than at microwave bands [37], [112]. The penetration losses are typically larger at higher frequencies. Hence, it is difficult to cover inside with mmWave nodes deployed outside and vice-versa due to high penetration loss.

**High power consumption**: In addition to the challenges imposed by high pathloss, (1) shows that the transmit power needs to increase with the bandwidth if the signal-to-noise ratio (SNR) should remain intact [14]. Alternatively, directive antennas or MIMO technology can be used to direct the signal power spatially, which leads to an array gain and latter also provides the flexibility of spatial multiplexing [36]. MIMO/beamforming is considered essential for mmWave communications, particularly since the short wavelength at mmWave frequencies makes it possible to fit many half-wavelength spaced antennas into a small area. MIMO arrays are normally fully digital for sub 6 GHz systems, where each antenna requires a dedicated radio-frequency (RF) chain, including power amplifier (PA), low noise amplifier (LNA), data converter (ADC/DACs), mixer, etc. However, realizing a fully digital MIMO implementation at mmWave frequencies is a non-trivial task, using current circuit design technology [14]. Having hundreds (or even thousands) of antennas, each supported by a separate RF chain, requires a very compact circuit implementation. Moreover, due to the high bandwidth, the PAs and data converters are expensive and power consuming. Hence, it appears that fully digital mmWave MIMO implementations are currently infeasible from a cost-efficiency perspective. This is likely to change in the future, but, in the meantime, alternative low RF-complexity architectures have received much attention from the research community. In particular, hybrid analog/digital architectures are being considered, where the corresponding signal processing techniques must be redesigned to enable channel estimation and a good tradeoff between the spectral efficiency and energy consumption/hardware cost [37].

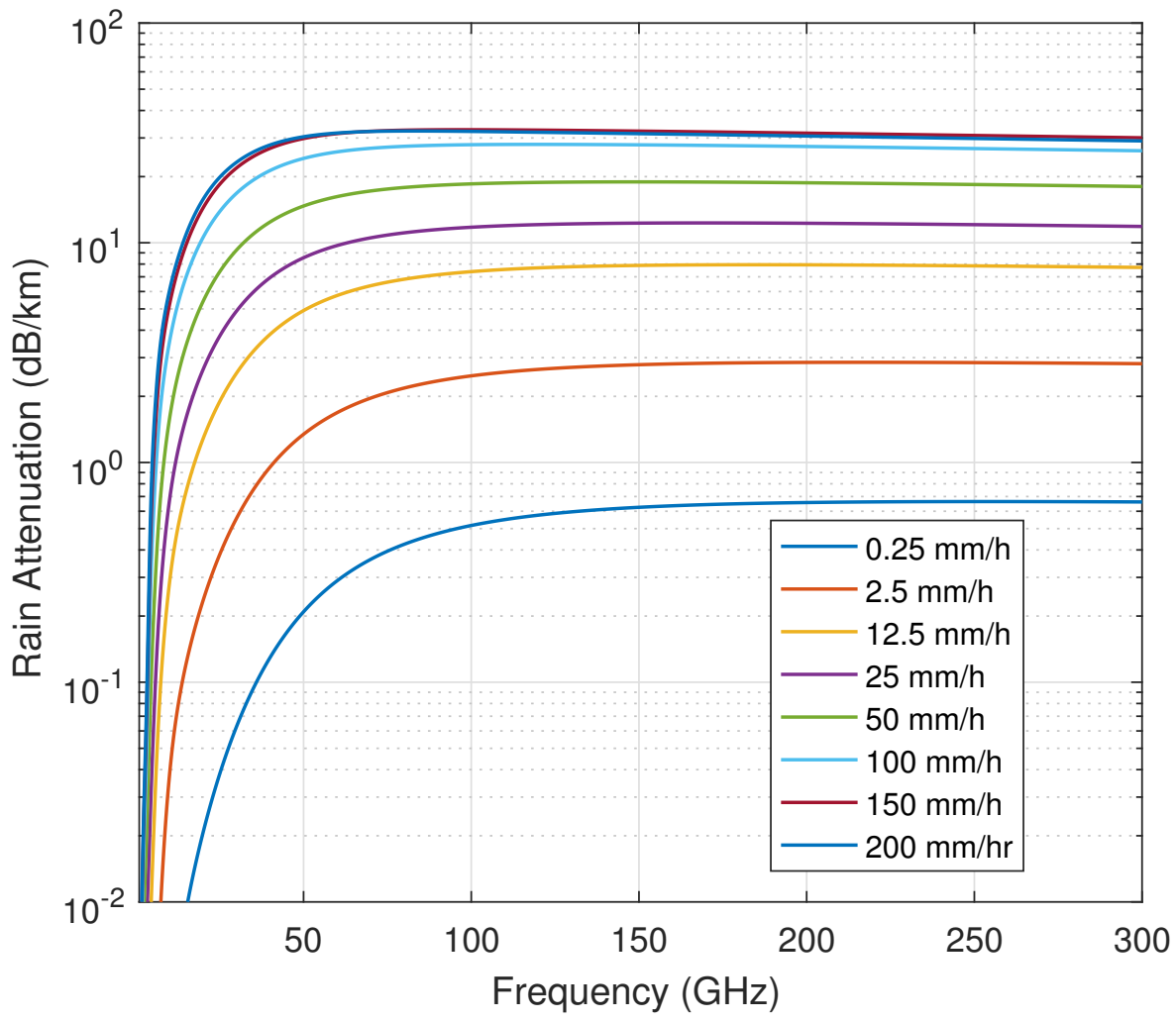

Fig. 3. Rain attenuation in different frequency bands [35].

**Narrow beamwidth and side-lobes**: To increase the transmission distance for mmWave, an array gain can be obtained by using directional antennas, MIMO, and beamforming. Consequently, the beamwidth of mmWave signals is normally narrow. When modeling the directivity, the radiation patterns are usually modeled in an idealized fashion, e.g., a constant large antenna gain within the narrow main-lobe and zero elsewhere. This idealized radiation pattern, often referred as the "flat-top model", was used in [22], [38]–[40] for system-level performance analysis. However, in practice, the radiation patterns are more complicated and implementation-dependent; the main-lobe gain is not constant and the side-lobe radiation is non-zero. The effect of side-lobe radiation and the gradual reduction of main-lobe gain caused by beam misalignment cannot be ignored. The maximum beamforming gain, which can be achieved only if the main-lobes of the transmitter and receiver are perfectly aligned, is rare due to practical implementation constraints. In 3GPP, a more practical two-dimensional directional antenna pattern [41] is adopted, where the antenna gain $G(\theta)$, with respect to the relative angle $\theta$ to its boresight, is given by

$$G(\theta) = \begin{cases} G_m 10^{-\frac{3}{10}\left(\frac{2\theta}{\omega}\right)^2}, & |\theta| \le \dfrac{\theta_m}{2}, \\ G_s, & \dfrac{\theta_m}{2} \le |\theta| \le \pi, \end{cases} \tag{3}$$

where $\omega$ denotes the half-power (3 dB) beamwidth and $\theta_m$ is the main-lobe beamwidth. $G_m$ and $G_s$ represent the maximum main-lobe gain and averaged side-lobe gain, respectively. Eq. (3) models how the signal changes with the beamwidth and receiving angles. The narrow beamwidth has a two-fold impact. While the pros will be discussed later, for the cons, narrow beamwidth leads to higher sensitivity to misalignment between the transmitter and the receiver, especially in mobile networks that should support high mobility. The reason for beam misalignment can be coarsely divided into two categories: 1) Imperfection of existing antenna and beamforming techniques [42]–[44], such as the analog beamforming impairments, array perturbations, oscillator locking-range based phase error, and the direction-of-arrival (DoA) estimation errors. Moreover, limited feedback may also cause the transmitter having only partial channel information and thus beaming misalignment [45]. 2) Mobility of communication UE [46], [47], which invokes tracking error and system reaction delay.

**Hardware impairments and design challenges**: In addition to above challenges, practical transceiver hardware are impaired by phase noise (PN), non-linear PAs, I/Q imbalance, and limited ADC resolution [48]. These effects limit the channel capacity [49], particularly when high spectral efficiency is envisioned. On the other hand, it was proved in [50] that MIMO communication links are less affected by hardware impairments than single-antenna links.

In mmWave communication systems, mixers are applied for signal up-conversion at the transmitter and down-conversion at the receiver using local oscillators to generate carrier signals operating at the desired carrier frequency. However, due to the random deviation of the output signal frequency around the carrier, it is infeasible that both oscillators at the transmitter and the receiver operate exactly at the same carrier frequency. Such a mismatch can be described by PN since the frequency offset yields a random phase difference for the time domain samples. Due to the high carrier frequency, mmWave communication systems are more sensitive to PN than conventional ones.

Another important hardware impairment in mmWave is the nonlinear PAs, since it is challenging to provide linear amplification to a signal with very wide bandwidth. In practice, each amplifier has a non-linear behavior, e.g., input signals of large amplitude are clipped and different frequencies are amplified differently. For the modeling of such non-linear characteristics, the modified Rapp model [51], is commonly used to describe the input-output relationship. Denoting $V_i$ the input voltage level of the PA, the output voltage level $V_o$ is described by the amplitude modulation to amplitude modulation (AM-AM),

$$V_o = \frac{rV_i}{(1 + (|rV_i|/V_s)^{2p})^{\frac{1}{2p}}}, \tag{4}$$

where $r$ is the small signal gain, $V_s$ is the limiting output amplitude, and $p$ controls the smoothness of the transition from linear operation to saturated operation. There are also other models, such as the Saleh model [52].

Moreover, the higher and larger frequency bands of mmWave communication systems cause many technical challenges in the design of circuit components and antennas. In [53], authors discussed in depth the challenges against the design of mmWave CMOS Radios, ranging from the device-level challenges to the architecture-level challenges. In particular, phase noise and IQ imbalance also cause severe technical challenges against realizing mmWave RF circuits [54]–[56]. In [57], authors have discussed and summarized the latest research progress on integrated circuits for mmWave communication systems. Authors have specifically summarized in details a few key technologies, including RF power amplifiers , mixers, high speed analog-to digital converters, on-chip and in-package antennas and 60 GHz voltage-controlled oscillators.

### B. Technical Potentials

Having listed the main challenges, we should not forget the main reasons for using mmWave communications.

**Large continuous unused bandwidth**: Compared to microwave communications, one of the major benefits of mmWave communications is the availability of large bandwidth, though wider bandwidth does not always lead to higher rates in the noise-limited region [8]. Currently, the available bandwidth for mobile networks (2G, 3G, 4G and LTE-Advanced spectrum) is globally smaller than 780 MHz and each major wireless provider has only a total of about 200 MHz spectrum [7]. This bandwidth is not sufficient for providing rates of Gbps to multiple devices, since a huge per-device spectral efficiency would be required. However, in mmWave bands, there are large chunks of bandwidth available for future mobile networks. As shown in Fig. 4, in the mmWave bands, the potentially available bandwidth can be more than 150 GHz [5], even excluding unfavorable bands such as the 60 GHz oxygen absorption band (57–64 GHz) and the water vapor ($H_2O$) absorption band (164–200 GHz). With 150 GHz of spectrum, a low spectral efficiency of 1 b/s/Hz is sufficient to deliver a rate of 150 Gbps. Having a low spectral efficiency simplifies the implementation and makes the end performance less affected by hardware impairments. The large unused frequency bands have therefore already attracted lots of interest. For instance, in October 2003, the FCC announced that the 71–76 GHz, 81–86 GHz, and 92–95 GHz frequency bands (collectively referred as the E-band) will be available for ultra-high-speed data communication including point-to-point WLAN, mobile backhaul, and broadband Internet access. A total of 12.9 GHz bandwidth is available in the E-band (60-90 GHz). More recently, in July 2016, FCC dedicated large bandwidths in mmWave bands for future cut-edge wireless communications, namely, 64-71 GHz unlicensed bands (plus previous 57-64 GHz) and 27.5-28.35 and 37-40 GHz licensed bands.

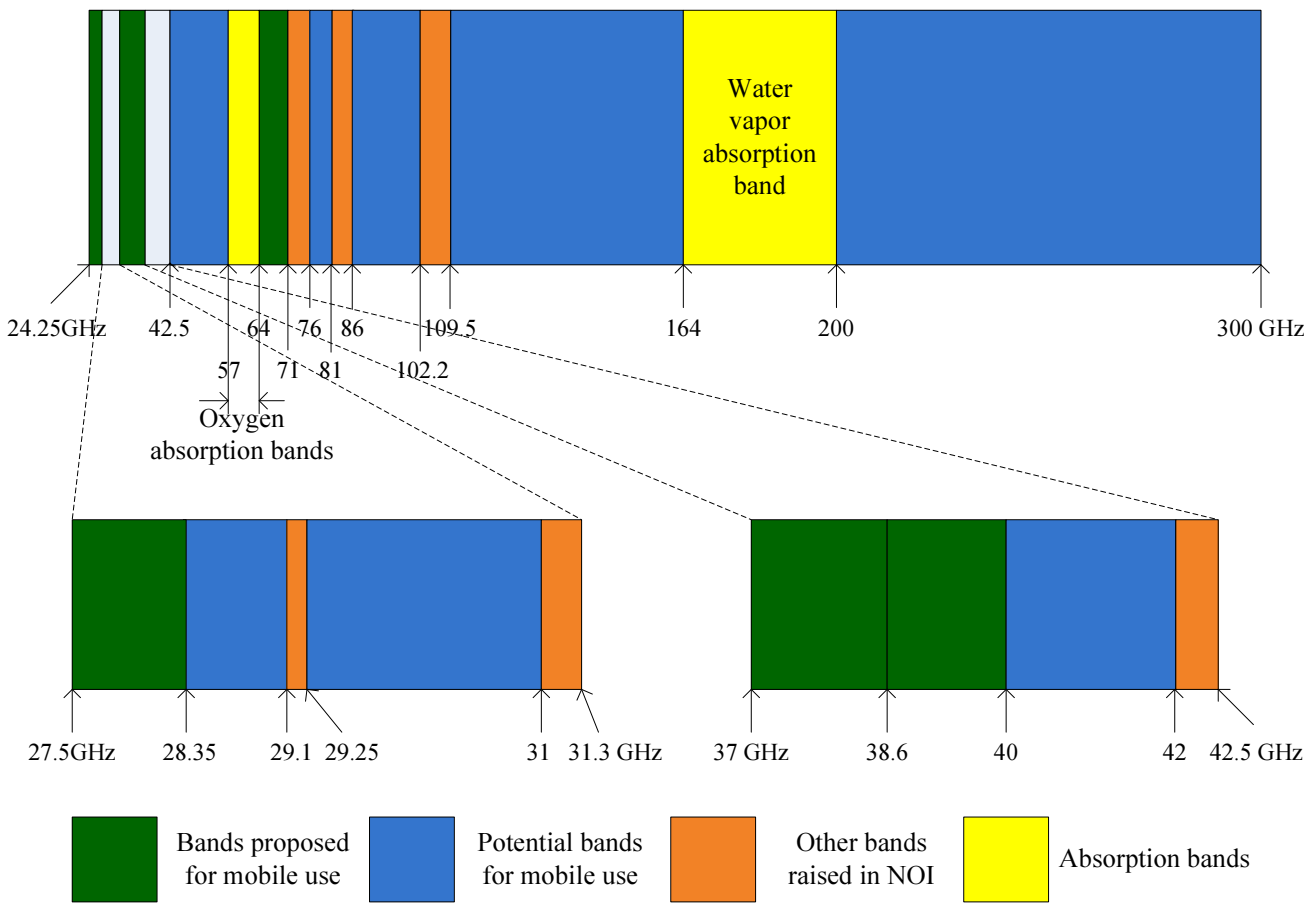

Fig. 4. Spectrum usage in mmWave bands.

**Short wavelength and narrow beamwidth**: Contrary to signals at sub-6 GHz bands, the mmWave signal has much shorter wavelength, which facilitates packing a large number of antennas into an array of compact size [6], [7]. This greatly expands the application range of large-scale antenna communications in the future mobile networks [9], [10]. At the same time, when having many antenna elements, the beamwidth is narrow [11]. The positive side of this property is the higher security against eavesdropping and jamming, and a larger resilience against co-user interference. This implies that the spectrum can be reused frequently in space, so many interfering point-to-point MIMO systems (or multiuser MIMO systems) can be deployed in a limited spatial region.

## III. Channel Measurements and Modeling

### A. Millimeter Wave Measurement Campaigns

Since the wavelength of mmWave bands is far shorter than in microwave bands below 6 GHz, the parameters for radio channel models will be quite different. Thus, understanding the mmWave propagation characteristics is the first task to design and develop mmWave communication systems. In general, parameters such as pathloss, delay spread, shadowing, and angular spread are used to characterize the radio propagation, which can be obtained through analyzing the data collected by various channel measurement campaigns in different environments. Extensive channel measurement campaigns

TABLE I
SUMMARY OF MMWAVE MEASUREMENT CAMPAIGNS

| Frequency (GHz) | Scenario | Site | Parameters | Ref. |
|---|---|---|---|---|
| 10 | Street canyon | Berlin | Large-scale parameters, time variance, and frequency dependence | [58] |
| | Outdoor to indoor | Belfort | Penetration losses, delay spread, and time variation | [58] |
| 11 | Urban outdoor | Ishigaki | Pathloss, RMS delay spread, shadow, and power delay profile | [59] |
| 15 | Street canyon | Helsinki | Pathloss, delay spread, and angular spread | [58] |
| | Street canyon | Stockholm | Pathloss and delay spread | [58] |
| | Office | | Pathloss, directional spread, and delay spread | [58] |
| | Airport | Helsinki | Pathloss, delay spread, and angular spread | [58] |
| | Outdoor to indoor | Stockholm | Penetration loss and RMS delay spread | [58] |
| 16 | Urban outdoor | Tokyo | Pathloss | [60] |
| 17 | Outdoor to indoor | Belfort | Penetration losses, delay spread, and time variation | [58] |
| 26 | Urban steet and Sidewalk | Tokyo | Pathloss | [61] |
| 28 | Street canyon | Helsinki | Delay spread, angular spread, and pathloss | [58] |
| | Street canyon | Berlin | Large-scale parameters, time variance, and frequency dependence | [58] |
| | Open square | Helsinki | Pathloss, delay spread, and angular spread | [58] |
| | Airport | Helsinki | Pathloss, delay spread, and angular spread | [58] |
| | Indoor (lab) and outdoor urban | Dallas | Penetration loss, pathloss, reflectivity, and AoA | [62] |
| | Dense urban | New York | Pathloss, RMS delay spread, shadow, PDP, AoA, and AoD | [63]–[65] |
| | Indoor and outdoor to indoor | New York | Penetration loss and reflection coefficients | [66] |
| | Dense urban | New York | Outage | [67] |
| | Outdoor to indoor | Göteborg | Excess loss, received power | [68] |
| | Office | New York | Pathloss, RMS delay spread, and shadow | [69], [70] |
| | Laboratory | | RMS delay spread and PDP | [71] |
| | Station and airport | Seoul | Pathloss and shadow | [72] |
| | Urban | Daejeon | Pathloss, RMS delay spread, angle spread, AoA, and AoD | [73], [74] |
| 37 | Urban steet and Sidewalk | Tokyo | Pathloss | [61] |
| 38 | Urban (Campus) | Austin | Pathloss, RMS delay spread, and PDP | [75] |
| | Urban (Campus) | Austin | Outage statistics | [76] |
| | Urban (Campus) | Austin | Pathloss, RMS delay spread, shadow, PDP, AOA, outage statistics | [77] |
| 40 | Indoor (lab) | | Penetration loss, reflectivity | [62] |
| 41 | Street canyon | Berlin | Large-scale parameters, time variance, and frequency dependence | [58] |
| 55 | Urban (street) | London | Pathloss, fading envelope, coherence bandwidth | [78] |
| 57 | Urban (street) | Oslo | Mean delay, delay spread, delay interval and delay window | [79] |
| 60 | Street canyon | Helsinki | Delay spread, angular spread, and pathloss | [58] |
| | Street canyon | Stockholm | Pathloss and delay spread | [58] |
| | Office | | Pathloss, directional spread, and delay spread | [58] |
| | Shopping mall | University of Bristol | Direction | [58] |
| | Outdoor to indoor | Stockholm | Penetration loss and RMS delay spread | [58] |
| | Indoor (lab) | | Penetration loss | [62] |
| | Urban (Campus) | Austin | Pathloss, RMS delay spread, and PDP | [75] |
| | Hospital | Japan | Pathloss, delay spread, and PDP | [80] |
| | Outdoor courtyard and in vehicle | | Pathloss, RMS delay spread, PDP, and AoA | [81] |
| | Street canyon | Berlin | Pathloss | [82] |
| | Street canyon | Berlin | Pathloss and delay spread | [83] |
| | Urban | Aachen | Angular, RSS | [84] |
| 72 | Office | New York | Penetration loss, delay spread, PDP | [85] |
| 73 | Dense urban | New York | Pathloss, delay spread, shadow, PDP, AoA, and AoD | [65], [86] |
| | Dense urban | New York | Outage | [67] |
| | Office | New York | Pathloss, RMS delay spread and shadow | [69], [70] |
| 81-86 | Roof-to-street and street canyon | Otaniemi and Kaisaniemi (Helsinki) | Frequency response, impulse responses, and delay | [87] |
| 82 | Street canyon | Berlin | Large-scale parameters, time variance, and frequency dependence | [58] |
| | Office | CEA-Leti | Delay and angular spread | [58] |
| 10 | Indoor hotspot | Oulu | Large-scale parameters, pathloss model | [88] |
| 11, 16, 28, 38 | Indoor | Jinan, China | Power delay, azimuth, elevation profile | [89] |
| 73 | rural | Virginia | pathloss model | [90] |
| 26 | Indoor | BeiJing | pathloss, shadow fading, and coherence bandwidth | [91] |

TABLE II
SUMMARY OF MMWAVE MEASUREMENT RESULTS ON DIRECTIONAL (D) AND OMNIDIRECTIONAL (O) PATHLOSS, DELAY SPREAD AND SHADOWING

| Freq. (GHz) | Environment | Scenario | Site | Tx/Rx Antenna (Height (m) etc.) | PL exp. | Delay spread (ns) | Shadowing (dB) | Ref. |
|---|---|---|---|---|---|---|---|---|
| 11 | Macro cellular | LoS/NLoS | Ishigaki | (O), Tx: 28, Rx: 3 | 2.6/3.4 | | 4.4/6.7 | [59] |
| 11 | Micro cellular | LoS | Ishigaki | (O), Tx: 8, Rx: 3 | 2.7 | | 2.5 | [59] |
| 11 | Outdoor hotspot | LoS | Ishigaki | (O), Tx: 3, Rx: 3 | 2.2 | | 2.5 | [59] |
| 28 | Outdoor cellular | LoS/NLoS | New York (Manhattan) | (D), 24.5-dBi Tx/Rx | 2.55/5.76 | | 8.66/9.02 | [7], [63] |
| 28 | Dense urban | LoS/NLoS | New York (Manhattan) | (O), Tx:7; 17, Rx: 1.5 | 2.1/3.4 | | 3.6/9.7 | [65] |
| 28 | Office | LoS/NLoS | New York (Brooklyn) | (D), V-V, 15-dBi, Tx: 2.5, Rx: 1.5 | 1.7/4.5 | 4.1/18.4 | 2.6/11.6 | [69] |
| 28 | Office | LoS/NLoS | New York (Brooklyn) | (O), V-V, 15-dBi, Tx: 2.5, Rx: 1.5 | 1.1/2.7 | | 1.7/9.6 | [69] |
| 28 | Office | LoS/NLoS | New York (Brooklyn) | (D), V-H, 15-dBi, Tx: 2.5, Rx: 1.5 | 4.1/5.1 | 12.8/18.7 | 8.0/10.9 | [69] |
| 28 | Office | LoS/NLoS | New York (Brooklyn) | (O), V-H, 15-dBi, Tx: 2.5, Rx: 1.5 | 2.5/3.6 | | 3.0/9.4 | [69] |
| 28 | Station | LoS/NLoS | Seoul | (D), Tx: 8, Rx: 1.5 | 2.15/4.06 | | 1.19/10.67 | [72] |
| 28 | Airport | LoS/NLoS | Seoul | (D), Tx: 8, Rx: 1.5 | 2.17/3.55 | | 1.33/7.61 | [72] |
| 28 | Urban | NLoS | Daejeon | (O), 24.5-dBi Tx:15 | 3.53 | 22.29 | 6.69 | [73] |
| 28 | Urban street | LoS/NLoS | Daejeon | (O), Tx:15, Rx: 1.6 | 1.90/3.15 | | 0.63/22.09 | [74] |
| 28 | Urban street | LoS/NLoS | New York (Manhattan) | (O), Tx:15, Rx: 1.6 | 1.81/3.03 | | 2.05/17.99 | [74] |
| 28 | Urban street | LoS/NLoS | New York (Manhattan) | (O), Tx:15, Rx: 1.6 | 1.87/2.97 | | 1.74/15.92 | [74] |
| 38 | Outdoor Cellular | LoS/NLoS | Austin | (D), 25-dBi TX and 13.3-dBi RX | 2.13/2.54 | 10.1 | 8.14/7.74 | [77] |
| 38 | Outdoor Cellular | LoS/NLoS | Austin | (D), 25-dBi TX and 13.3-dBi RX | 2.16/2.52 | 17.3 | 8.78/7.82 | [77] |
| 38 | Outdoor Cellular | LoS/NLoS | Austin | (D), 25-dBi TX and 13.3-dBi RX | 2.03/2.40 | 4.8 | 5.31/5.27 | [77] |
| 38 | Outdoor Cellular | LoS/NLoS | Austin | (D), 25-dBi TX and 13.3-dBi RX | 2.74/2.97 | 17.5 | 12.46/11.16 | [77] |
| 38 | Outdoor Cellular | LoS/NLoS | Austin | (D), 25-dBi TX and 25-dBi RX | 2.25/3.29 | 13.5 | 6.51/11.63 | [77] |
| 38 | Outdoor Cellular | LoS/NLoS | Austin | (D), 25-dBi TX and 25-dBi RX | 2.38/3.20 | 15.1 | 14.12/8.97 | [77] |
| 38 | Outdoor Cellular | LoS/NLoS | Austin | (D), 25-dBi TX and 25-dBi RX | 2.01/2.70 | 5.3 | 6.56/5.54 | [77] |
| 38 | Outdoor Cellular | LoS/NLoS | Austin | (D), 25-dBi TX and 25-dBi RX | 2.99/4.17 | 16.5 | 13.92/8.90 | [77] |
| 55 | Urban (street) | LoS/NLoS | London | (D), Tx: 10, Rx: 2 | 3.6/10.4 | | | [78] |
| 60 | Hospital | LoS | Japan | Tx (O): 1.31, Rx (D): 2.04 | 1.34 | 6.7 | | [80] |
| 60 | Hospital | LoS | Japan | (O), Tx : 2.30, Rx: 2.04 | 1.86 | 6.3 | | [80] |
| 60 | Hospital | LoS | Japan | (O), Tx: 0.75, Rx: 1.58/2.29 | 2.23 | 3.8 | | [80] |
| 60 | Street canyon | LoS/NLoS | Berlin | (O) | 2.02/3.06 | | | [83] |
| 72 | Indoor | NLoS | New York (Brooklyn) | 20-dBi Tx/Rx: 1.61 | | | | [85] |
| 73 | Office | LoS/NLoS | New York (Brooklyn) | (D), V-V, 15-dBi, Tx: 2.5, Rx: 1.5 | 1.7/5.3 | 3.3/13.3 | 2.1/15.6 | [69] |
| 73 | Office | LoS/NLoS | New York (Brooklyn) | (O), V-V, 15-dBi, Tx: 2.5, Rx: 1.5 | 1.3/3.2 | | 1.9/11.3 | [69] |
| 73 | Office | LoS/NLoS | New York (Brooklyn) | (D), V-H, 15-dBi, Tx: 2.5, Rx: 1.5 | 4.7/6.4 | 21.2/10.3 | 9.0/15.8 | [69] |
| 73 | Office | LoS/NLoS | New York (Brooklyn) | (O), V-H, 15-dBi, Tx: 2.5, Rx: 1.5 | 3.5/4.6 | | 6.3/9.7 | [69] |
| 73 | Outdoor cellular | LoS/NLoS | New York | (D), Tx: 7, Rx: 4 | 2.68/4.72 | | 6.79/9.52 | [86] |
| 73 | Outdoor cellular | LoS/NLoS | New York | (D), Tx: 7, Rx: 2 | 2.77/4.64 | | 5.33/8.32 | [86] |
| 73 | Outdoor cellular | LoS/NLoS | New York | (D), Tx: 17, Rx: 4 | 2.52/3.90 | | 4.00/10.81 | [86] |
| 73 | Outdoor cellular | LoS/NLoS | New York | (D), Tx: 17, Rx: 2 | 2.32/3.76 | | 2.71/8.81 | [86] |
| 73 | Dense urban | LoS/NLoS | New York (Manhattan) | (O), Tx:7; 17, Rx: 2 | 2.0/3.3 | | 5.2/7.6 | [65] |
| 73 | Dense urban | LoS/NLoS | New York (Manhattan) | (O), Tx:7; 17, Rx: 4.06 | 2.0/3.5 | | 4.2/7.9 | [65] |
| 10 | Indoor | LoS/NLoS | Oulu | Tx: 2, Rx: 1.6 | 1.4/3.3 | 17.4/30.4 | 1.4/2.6 | [88] |
| 11/16/28/38 | Indoor | | JiNan | Tx: 2.6, Rx: 1.45 | | 9.1/10.2/10.6/8.5 | | [89] |
| 73 | rural | LoS/NLoS | Virginia | Tx:110, Rx: 1.6-2 | 2.16/2.75 | | 1.7/6.7 | [90] |
| 26 | Hall | LoS | BeiJing | Tx: 2.5, Rx: 2 | | 16.08 | 0.098 | [91] |

[58]–[87], covering potential mmWave mobile communication bands, like 10, 28, 38, 60, and 82 GHz, have already been performed. Table I summarizes the performed measurement campaigns, where in particular NYU WIRELESS and the EU mmMAGIC project make contributions.[1] The researchers from NYU WIRELESS have performed many measurements at 28, 38, 60, 72 and 73 GHz since year 2011, and obtained abundant measurement results. Based on various measurement results (including those from NYU WIRELESS) and ray tracing analysis, 3GPP [21], [99] also provides new modeling features like: i) dynamic LoS/non-LoS (NLoS) blockage [2], spatial consistency and penetration modeling, ii) extension of power delay/angle profiles, and iii) the pathloss model based on one-meter reference distance. The EU mmMAGIC project plans more than 60 single-frequency measurement campaigns, covering eight frequency bands from 6 to 100 GHz, under typical environments and scenarios, including urban micro-cellular (UMi) (street canyon, open square), indoor (office, shopping mall, airport), outdoor-to-indoor (O2I) and two scenarios with very high user densities (stadium and metro station) [58].

*1) Pathloss and Shadowing:* Pathloss and shadowing are the two most important large-scale characteristics of the radio channel, which have been reported for various environments both in LoS and NLoS cases. It is common to determine the pathloss exponent that best fits the measurements to Friis transmission formula in (2), while the shadowing parameters are used to model the random deviations from this model. Table II summarizes the main mmWave measurement results for directional and omnidirectional pathloss and shadowing.

*2) Power Delay Profile and Delay Spread:* The shape of power delay profile (PDP) in measurements is a single or superposition of multiple exponentially decaying spectrums. The delay spread denotes an extent of the multipath power spread over the PDP, which is an important parameter that determines the inter-symbol interference in single-carrier transmission and the frequency-flatness of the subcarriers in multi-carrier transmission. In Table II, we also summarize the delay spread of various measurement results.

### B. Millimeter Wave Channel Modeling

Channel models are important for evaluating and analyzing the performance in system-level simulations. For mmWave, some recent channel modeling results are reported in [92]–[98]. For more general mobile and wireless networks, several channel models have proposed continuously, which are illustrated in Fig. 5. To give a historical perspective, we give a brief introduction as follows (even though some models are not for mmWave).

*1) 3GPP Spatial Channel Model and SCM-Extended:* The 3GPP Spatial Channel Model (SCM) [41] was proposed in 2003 and supports six delay paths with 5 MHz bandwidth in the 2 GHz frequency band under three scenarios: Suburban Macro (SMa), Urban Macro (UMa), and UMi. The SCM-Extended model [99] further extended the SCM, by supporting bandwidths of up to 100 MHz.

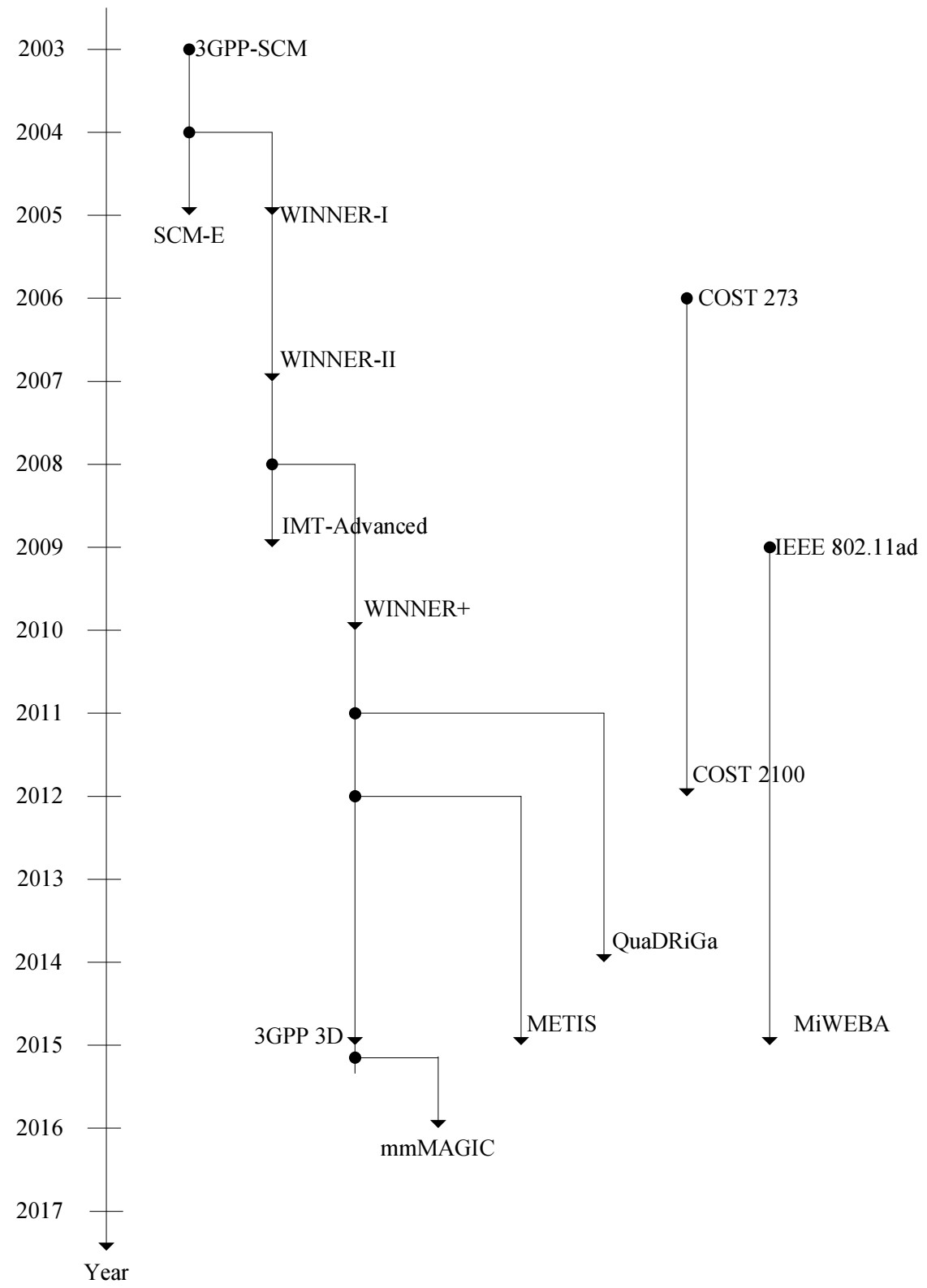

Fig. 5. Available channel models for mobile and wireless networks [58].

*2) WINNER I/II/+ Model:* As the EU flagship mobile technological projects for 4G, the WINNER I/II/+ projects developed several channel models for mobile networks. The WINNER I model [100] is an antenna-independent model, in which different antenna configurations and different element patterns can be used. In order to support more scenarios, such as outdoor-to-indoor and indoor-to-outdoor, and elevation in indoor scenarios, the WINNER II model [101] was developed. It includes scenario-dependent polarization modeling, and thus improves the accuracy for cross-polarized MIMO antennas. The parameter tables were reviewed and additional measurements were done in order to cover the complete 1–6 GHz frequency range. As a major upgrade of the WINNER II model, the WINNER+ model [102] supports 3D propagation effects. WINNER II and WINNER+ models are also antenna independent since they are based on double-directional channel representations.

*3) 3GPP 3D Model [103], [104]:* This model is defined in the 2 GHz band at a relatively small bandwidth of 10 MHz. It has consolidated parameters for the two most commonly used scenarios: UMa and UMi, which are further split into LoS, NLoS and O2I propagation. The core part of this model, i.e., the small-scale fading (SSF) model, is identical to the WINNER+ model. Thus, the same parameters can be used and similar functionality is provided. More recently, 3GPP TR38.900 [98] summarizes the recent channel modeling re-sults of 3GPP for the band above 6 GHz and up to 100 GHz. The models in [98] are comprehensive including street canyon, open area, rooftop, indoor, backhaul, D2D/V2V and stadium etc.

[1]Note that there are a lot of excellent measurement campaigns and results. For space limits, we can only list a part of the existed literatures here. The results, especially before 2012, are mostly omitted.

[2]COST 259 also include the LoS/NLoS blockage probability model.

*4) COST 273/2100 Model:* The COST 273 model [105] was an evolution of the earlier COST 259 channel model towards mobile broadband by using MIMO technologies. The model is based on the concept of geometrically located multipath clusters in 2D propagation environment to model the interrelationship between Angle-of-Arrivals (AoAs) and Angle-of-Departures (AoDs). This concept is effective to keep spatial consistency, and can evaluate the performance of MIMO beamforming and multi-cell transmission more accurately. As an evolution of the COST 273 model, the COST 2100 model [106] used visibility regions (VRs) introduced in COST 259 to model the scenario-variation. These VRs make the evaluation of multi-cell and heterogeneous transmissions more practical by considering the VR of BSs from each UE. This model also extends the multipath clusters to 3D propagation environments.

*5) QuaDRiGa Model [107]:* As an open source implementation of the 3GPP-3D channel model, the QuaDRiGa channel model is further extended with the features of spatial consistency (to accurately evaluate the performance of massive MIMO) and multi-cell transmissions by exploiting the approach in SCM-E and COST 273.

*6) IEEE 802.11ad Model [108]:* This model was developed in 2010 to support indoor short-range communications, such as in offices and homes using 60 GHz unlicensed band. The model is quasi-deterministic (QD): specular components, such as the LoS path and single and double bounce reflections are modeled deterministically in 3D propagation environments, while other contributions are modeled stochastically as random components in the cluster. One of the important features is the support for blockage.

*7) MiWEBA Model [109]:* The model is an extension of the IEEE802.11ad channel model towards outdoor access, backhaul/fronthaul, and device-to-device (D2D) scenarios. The approach is QD, where specular components are modeled deterministically, while other components are modeled stochastically. In this way, beamforming and path-blocking models can be supported. Furthermore, the first 60 GHz pathloss model in an UMi environment was developed in MiWEBA. The effect of the ground reflection paths traveling closely in space to the LoS path has been found to be of high significance.

*8) IMT-Advanced Model [110]:* This model consists of a primary module and an extension module (not specific to mmWave), where the former is based on the WINNER II channel model, while the latter enhances the support for variable BS antenna heights, street widths, and city structures.

*9) METIS Model [111]:* This model consists of a map-based (deterministic) model, a stochastic model, and a hybrid model as a combination of both. The stochastic model extends the WINNER+ and 3GPP-3D models to support 3D shadowing maps, mmWave parameters, direct sampling of the power angular spectrum, and frequency dependent pathloss models. Based on extensive measurement campaigns, lists of channel parameters for <6 GHz and 50 to 70 GHz bands are available.

*10) mmMAGIC Model [58]:* This is a statistical channel model for link- and system-level simulations that is designed for the entire frequency range from 6 to 100 GHz for a large variety of scenarios. The model uses the theoretical approach of the existing 3GPP 3D channel model. However, it extends it in the following ways: i) the spatial accuracy regarding path and sub-path distributions is substantially improved, ii) a realistic non-uniform distribution of sub-path amplitudes is included, iii) sub-paths can be modeled using spherical waves; iv) there is consistency over the frequency range (6 − 100 GHz), v) there is frequency-dependent antenna models; vi) providing continuous variations over time, vii) mmWave-specific random blockage, clustering and scattering objects are being modeled, and viii) the reflection from the ground or the floor is modeled.

*11) 3GPP-like 5G Model [22], [112]:* In [112], the outdoor model is established for the bands from 6 GHz to 100 GHz. An initial 3D channel model which includes: 1) typical deployment scenarios for urban microcells and urban macrocells, and 2) a baseline model for incorporating pathloss, shadow fading, LoS probability, penetration, and blockage models for the typical scenarios. Various processing methodologies such as clustering and antenna decoupling algorithms are also included in [112]. In [21], [22], the indoor model is established for office and shopping mall environments. The measurement results show that the smaller wavelengths introduce an increased sensitivity of the propagation models to the scale of the environment and show some frequency dependence of the pathloss as well as increased occurrence of blockage.

## IV. MIMO Design for mmWave Communications

While conventional microwave communications, operating below 6 GHz, can cover many users in wide coverage areas, mmWave communications mainly provide local-area coverage and thus feature fewer users. Moreover, there is limited scattering in outdoor mmWave communications, in contrast to the rich scattering in conventional microwave communications [7]. Due to these differences, mmWave MIMO systems have different constraints and requirements, requiring different transceiver designs. In this section, we first describe hardware architectures for mmWave MIMO systems, for the use in mobile networks. We discuss how to design the signal processing techniques, including channel estimation, channel-tracking, and precoding/combining for millimeter MIMO systems.

### *A. MIMO Architectures*

The conventional MIMO system is fully digital, where all the signal processing techniques are performed at baseband as shown in Fig. 6 (a) [36]. As explained in Section II, for mmWave communication with many antennas and high bandwidth, the conventional fully digital MIMO architecture requires many energy-intensive RF chains, leading to an unaffordable energy consumption and hardware cost. Therefore, although the fully digital architecture most likely will be available at some point in the future, alternative architectures are required for emerging mmWave mobile networks.

To reduce the implementation complexity, a fully analog architecture has been adopted in indoor mmWave communications, such as 60 GHz WLAN [114]. As shown in Fig. 6 (b), only one RF chain is employed to transmit a single data stream, and the analog circuit (e.g., realized by analog phase-shifters) is utilized to partially adjust the signals (e.g., phases of signals) to achieve an array gain. The advantage of the fully analog architecture is that it only requires one RF chain, leading to quite low hardware cost and energy consumption [114]. However, since the analog circuit can only partly adjust the signals, it is hard to adjust the beam to the channel conditions and this leads to a considerable performance loss [14], particularly for mobile users. In addition, fully analog architecture can only support single-stream transmission, which cannot achieve the multiplexing gain to improve spectral efficiency [14].

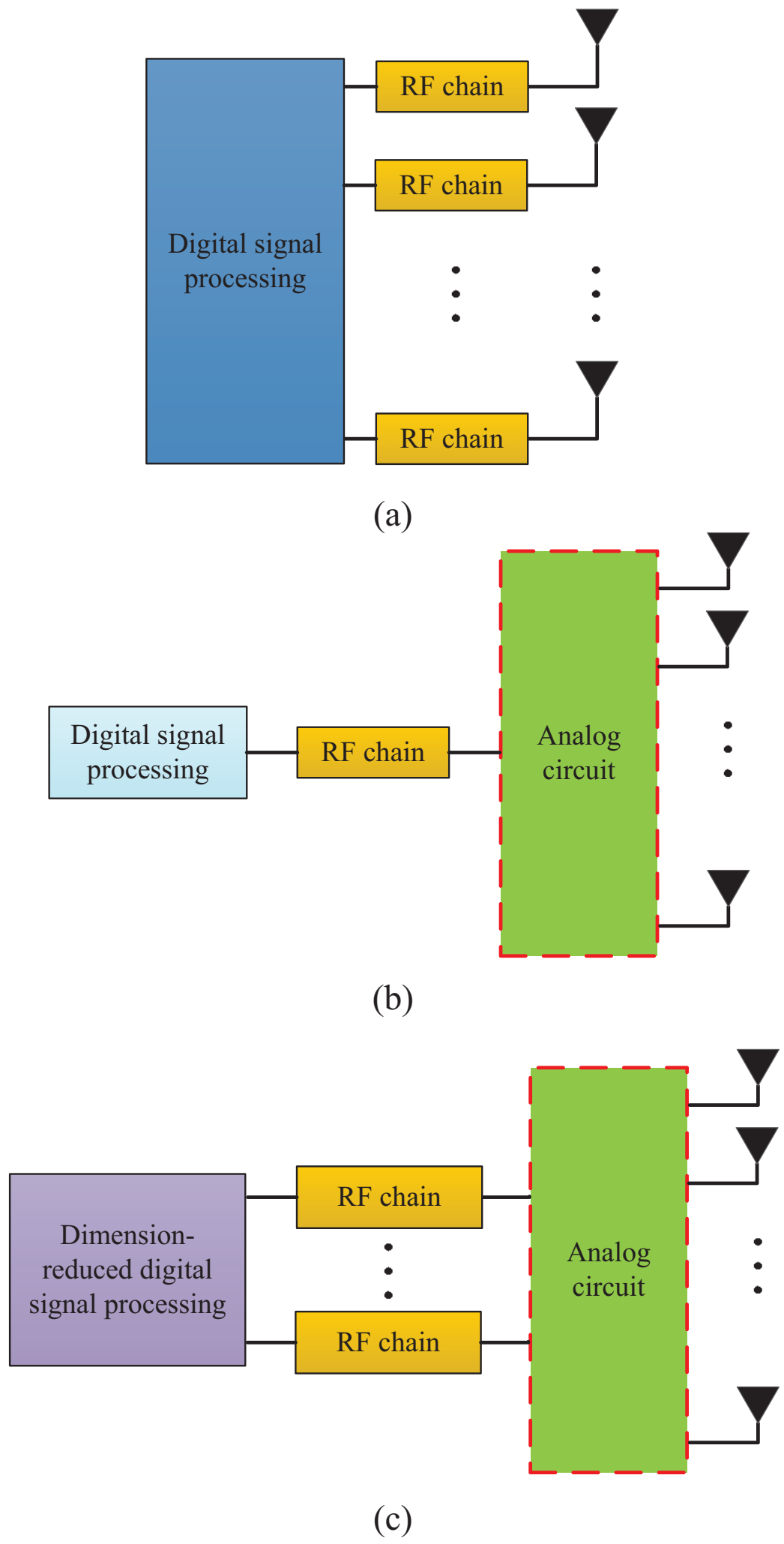

Fig. 6. MIMO architectures: (a) fully digital architecture; (b) fully analog architecture; (c) hybrid architecture.

New MIMO architectures need to be designed to balance between the benefits of fully digital and fully analog architectures. The hybrid analog-digital architecture is the key solution [122]–[124]; the hybrid architecture was agreed to be deployed in future 5G systems at the 3GPP RAN1 meeting in June 2016 [125], as described later. The hybrid architecture can be considered an extension of the fully analog architecture to the multi-stream scenario. As shown in Fig. 6 (c), the key idea is to divide the conventional digital signal processing (e.g., precoding and combining) of large size into two parts: a large-size analog signal processing (realized by analog circuits) and a dimension-reduced digital signal processing (requiring a small number of RF chains). Since there is often only a small number of effective scatterers at mmWave frequencies, each user has a MIMO channel matrix with low rank [7]. Hence, the optimal number of data streams is generally much smaller than the number of antennas. Since the number of streams determines the minimum required number of RF chains, this number can be significantly reduced by the hybrid architecture, leading to reduced cost and energy consumption.

The analog circuits of the hybrid architecture can be implemented by different circuit networks, leading to different hardware constraints [14], [115], [116]. Next, we describe the three typical implementation networks that are illustrated in Fig. 7. The choice of architecture affects not only the signal processing design but also the performance of mmWave MIMO systems.

*N1) Fully-connected network with phase-shifters:* In this network, each RF chain is connected to all antennas via phase-shifters, as shown in Fig. 7 (a) [117]. Hence, a highly directive signal can be achieved by adjusting the phases of transmitted signals on all antennas [117]. By employing such a network, all elements of the analog precoder/combiner have the same fixed amplitude.[3] Difficult factors in the implementation can be the addition of the analog signals at each antenna and selection of a collection of phase-shifts that are suitable over the entire bandwidth.

*N2) Sub-connected network with phase-shifters:* In this network, each RF chain is only connected to a subarray via phase-shifters, as shown in Fig. 7 (b) [113], [121]. Due to this limitation, for each RF chain, only the transmitted signals on a subset of antennas can be adjusted. Therefore, compared with N1, the achieved array gain and directivity is reduced proportionally to the number of subarrays [113]. However, this network might be preferred in practice, since there is no need to add analog signals at the antenna inputs and the number of phase-shifters required in this network is also significantly reduced. Moreover, it has been shown to achieve the performance close to that of N1 [121]. This network imposes two hardware constraints [121]: i) the analog precoder/combiner should be a block diagonal matrix; and ii) All the nonzero elements of the analog precoder/combiner have the same fixed amplitude.

*N3) Lens antenna array:* An alternative quite different from the networks discussed above is to utilize a lens antenna array, as shown in Fig. 7 (c) [126]. The lens antenna array (a feed antenna array placed beneath the lens) can realize the functions of signal emitting and phase-shifting simultaneously [126].

[3]Here we assume the resolution of phase-shifter is sufficiently high, so that the phase of transmitting signal can be arbitrarily adjusted. In practice, the phase-shifters with finite resolution will incur phase noise and degrade the performance of hybrid precoding and combing. In this case, how to design the signal processing techniques is an interesting topic of future research, and some initial works can be found in [118]–[120].

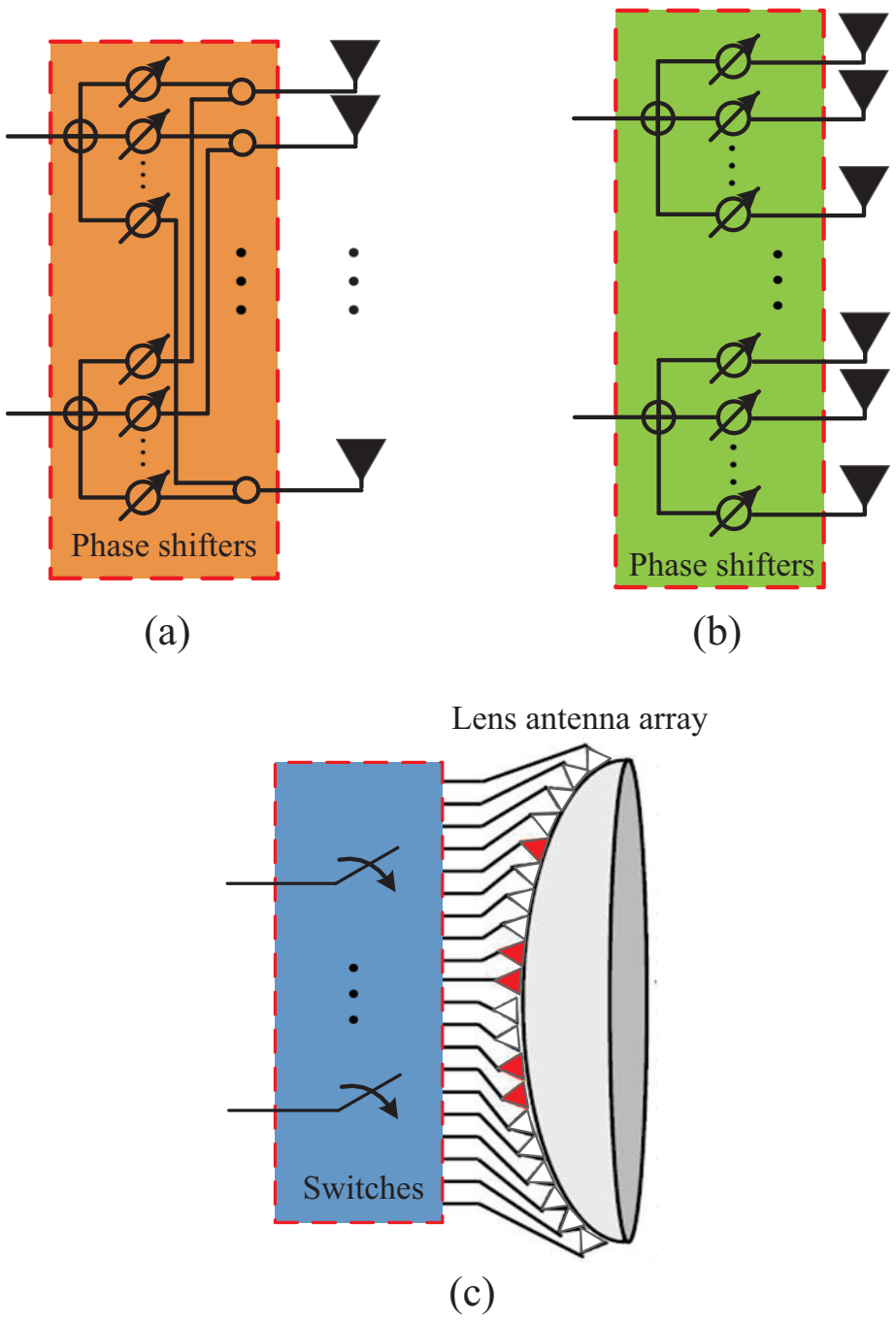

Fig. 7. Analog circuit with different networks: (a) Fully-connected network with phase-shifters (N1); (b) Sub-connected network with phase-shifters (N2); (c) Lens antenna array (N3).

It can concentrate the signals from different propagation directions (beams) on different feed antennas. As the scattering for outdoor mmWave communications is not rich [7], the number of effective propagation paths is usually limited, and the channel power will be concentrated on only a small number of beams. Therefore, the selecting network can be used to significantly reduce the MIMO dimension as well as the number of RF chains without major performance loss [126]. With careful design, the lens antenna array can excite several orthogonal beams spanning the whole space. If the directions of channel paths coincide with the directions of the orthogonal beams, an array gain similar to N1 can be achieved[4] [126]. Moreover, compared with phase-shifter network (including a large number of phase-shifters, power splitters/combiners, and signal/control lines) in N1, the hardware cost and energy consumption incurred by lens antenna array in N3 is relatively low [128], [251]. Essentially, the lens antenna array plays the role of a discrete Fourier transform (DFT).[5] Therefore, in this network, each column of the analog precoder/combiner restricts to a DFT column [126].

### *B. Channel Estimation with Hybrid Architecture*

Channel state information (CSI) is essential to benefit from the array gain provided by multiple antennas. Acquiring CSI is particularly challenging in mobile networks, where the channels can change rapidly. With a fully digital receiver, the use of pilot transmission is the most efficient way to acquire CSI [6], [130]. The channel estimation is more complicated in a hybrid mmWave architecture, since we cannot extract the actual received signals on all antennas simultaneously. We will now discuss alternative methods for acquiring CSI in a hybrid architecture.

[4]If not, power leakage will happen, leading to some performance loss after the selecting network [127].

[5]It is worth pointing out that mathematically, N3 is equivalent to the structure based on butter matrix proposed in [129], which aims to improve the performance of antenna selection for microwave MIMO with high correlation.

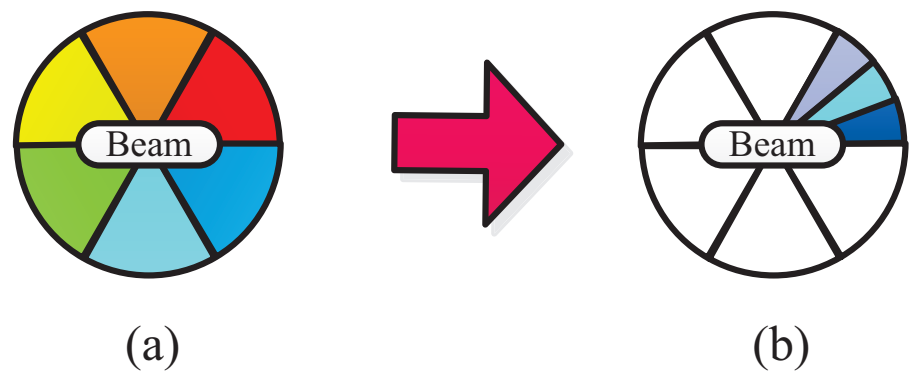

Fig. 8. Beam-training: (a) Wide beamwidth; (b) Narrow beamwidth.

To exemplify the signal processing, we consider a single-user multi-stream system. The transmitter employs $N_{\mathrm{T}}$ antennas and $N_{\mathrm{RF}}^{\mathrm{T}}$ RF chains, the receiver employs $N_{\mathrm{R}}$ antennas and $N_{\mathrm{RF}}^{\mathrm{R}}$ RF chains, and $N_{\mathrm{D}} \leq \min\left\{N_{\mathrm{RF}}^{\mathrm{T}}, N_{\mathrm{RF}}^{\mathrm{R}}\right\}$ parallel data streams are transmitted from the transmitter to the receiver. In this subsection, we mainly consider narrowband systems. Note that for broadband systems, the analog circuit is fixed for the whole bandwidth. As a result, the analog signal processing (e.g., precoding) cannot be adaptively adjusted according to different frequencies. This will lead to more challenges in signal processing design, which requires future investigation, but some pioneering works can be found in [137], [140]. The narrowband system model of the hybrid architecture can be presented as [117]

$$\mathbf{y} = \mathbf{W}^H \mathbf{H}\mathbf{F}\mathbf{s} + \mathbf{W}^H \mathbf{n}, \tag{5}$$

where $\mathbf{s}$ and $\mathbf{y}$ both of size $N_{\mathrm{D}} \times 1$ are the transmitted and received signal vectors, respectively, $\mathbf{H}$ of size $N_{\mathrm{R}} \times N_{\mathrm{T}}$ is the mmWave MIMO channel, which can be modeled as described in Section III, $\mathbf{n}$ of size $N_{\mathrm{R}} \times 1$ is the noise vector. $\mathbf{F} = \mathbf{F}_{\mathrm{A}}\mathbf{F}_{\mathrm{D}}$ of size $N_{\mathrm{T}} \times N_{\mathrm{D}}$ is the hybrid precoder, where $\mathbf{F}_{\mathrm{A}}$ of size $N_{\mathrm{T}} \times N_{\mathrm{RF}}^{\mathrm{T}}$ is the analog precoder and $\mathbf{F}_{\mathrm{D}}$ of size $N_{\mathrm{RF}}^{\mathrm{T}} \times N_{\mathrm{D}}$ is the digital precoder. Similarly, $\mathbf{W} = \mathbf{W}_{\mathrm{A}}\mathbf{W}_{\mathrm{D}}$ of size $N_{\mathrm{R}} \times N_{\mathrm{D}}$ is the hybrid combiner, where $\mathbf{W}_{\mathrm{A}}$ of size $N_{\mathrm{R}} \times N_{\mathrm{RF}}^{\mathrm{R}}$ is the analog combiner and $\mathbf{W}_{\mathrm{D}}$ of size $N_{\mathrm{RF}}^{\mathrm{R}} \times N_{\mathrm{D}}$ is the digital combiner.

The precoder and combiner matrices should be selected based on the current channel realization $\mathbf{H}$, but estimating this matrix is non-trivial [141]. Firstly, due to the lack of antenna gain before the establishment of the transmission link, the SNR for channel estimation can be quite low. Secondly, the number of RF chains in the hybrid architecture is usually much smaller than the number of antennas (i.e., $N_{\mathrm{RF}}^{\mathrm{T}} \ll N_{\mathrm{T}}$ and $N_{\mathrm{RF}}^{\mathrm{R}} \ll N_{\mathrm{R}}$), so we cannot simultaneously obtain the sampled signals on all receive antennas. As a result, the traditional channel estimation schemes [138], [139] requiring the sampled signals on all antennas will involve unaffordable pilot overhead in a hybrid architecture. To solve this problem, two dominant categories of channel estimation schemes have been proposed in the mmWave literature.

The key idea of the first category is to reduce the dimension of channel estimation problem, by dividing it into two steps.

In the first step, it performs the beam-training between the transmitter and receiver to obtain the analog precoder $\mathbf{F}_{\mathrm{A}}$ and analog combiner $\mathbf{W}_{\mathrm{A}}$. In the second step, the effective channel matrix $\mathbf{W}_{\mathrm{A}}^{H}\mathbf{H}\mathbf{F}_{\mathrm{A}}$ in the analog domain is estimated by classical algorithms, such as least squares (LS) [139]. Note that the size of $\mathbf{W}_{\mathrm{A}}^{H}\mathbf{H}\mathbf{F}_{\mathrm{A}}$ is $N_{\mathrm{D}} \times N_{\mathrm{D}}$, which is much smaller than that of the original channel matrix (i.e., $N_{\mathrm{D}} \ll N_{\mathrm{T}}, N_{\mathrm{R}}$). Therefore, the pilot overhead in the second step is relatively low, which is proportional to $N_{\mathrm{D}}$. The remaining difficulty lies in how to design the efficient beam-training scheme to find the optimal $\mathbf{F}_{\mathrm{A}}$ and $\mathbf{W}_{\mathrm{A}}$.

To achieve this goal, two primary approaches have been proposed. The first one is to extend the traditional single-beam training schemes standardized in IEEE 802.11ad/802.15.3c to multi-beam training [47], [142]. For example, the single-beam training scheme in IEEE 802.11ad consists of three phases [142]: i) Sector level sweep (SLS): the wide beamwidth is considered at first, as shown in Fig. 8 (a). The transmitter and receiver try all possible wide beam pairs. With the channel feedback to indicate the largest received SNR [143], [144], the best beam pair (according to some criterion) can be selected for the next phase; ii) Beam-refinement protocol (BRP): the narrow beamwidth is considered as shown in Fig. 8 (b), and the beams are trained in a similar way within the previous selected wide beam pair; iii) Beam tracking: a periodic beam-refinement is performed for the time-varying channels. Repeating such procedure above for each RF chain to select its corresponding beams, we can obtain the optimal $\mathbf{F}_{\mathrm{A}}$ and $\mathbf{W}_{\mathrm{A}}$. This hierarchical search could significantly reduce the training overhead, but its performance heavily depends on the designed training beam codebook. For the fully analog architecture, forming a wide beam usually requires the deactivation of some antennas and thus reduces the total transmit power. While for the hybrid architecture, it is possible to design wide beam without reducing the transmit power [145], [146]. Another scheme can be found in [164], where a novel codebook-based beam-training together with the following hybrid precoding design was proposed. Specifically, an RF codebook, denoted as $\mathcal{F}_{\mathcal{CB}}$, is used to specify the possible set of RF beamforming vectors by considering the practical limitation of phase-shifters. Then, the original hybrid precoding problem is transformed into a joint codeword selection and precoding design problem, which is essentially a group sparsity constrained optimization problem and only requires effective channels $\mathbf{H}\mathcal{F}_{\mathcal{CB}}$. Based on that, a beam-training procedure is proposed to obtain effective channels with less signaling feedback by utilizing the beam-domain sparse property of mmWave channels, and efficient algorithms are developed for maximizing both the spectral efficiency and the energy efficiency. The second approach is to employ algorithms developed from machine learning to realize the beam-training, since it can be considered as a combinatorial problem with a finite number of beams to be searched. For example, in [147], a tabu search (TS)-based beam-training scheme is proposed. It first selects an initial beam pair, and defines the neighbors of this pair (e.g., only one RF chain changes its beam and the others keep fixed). Then, the training procedure is only executed within the neighbors with much smaller size to find a pair which: i) enjoys the best performance; ii) is not the tabu beam pair according to some criteria. After a small number of iterations, it can obtain $\mathbf{F}_{\mathrm{A}}$ and $\mathbf{W}_{\mathrm{A}}$ with satisfying performance. Other algorithms, such as local search algorithm [148], can be also used. They can usually significantly reduce the pilot overhead, but the robustness cannot be guaranteed, since these algorithms are performed in a random way.

The second category of channel estimation schemes is to exploit the sparsity of mmWave MIMO channels. Instead of estimating the effective channel matrix $\mathbf{W}_{\mathrm{A}}^{H}\mathbf{H}\mathbf{F}_{\mathrm{A}}$ of small size, it can directly obtain the complete channel matrix $\mathbf{H}$ with low pilot overhead. To explain the basic idea, we consider the simplest case where both the transmitter and receiver use one RF chain for channel estimation[6] [141]. In time slot $m$, the transmitter uses a hybrid precoder $\mathbf{f}_m$ of size $N_{\mathrm{T}} \times 1$ to transmit the pilot $s_m$ to the receiver. The received pilot $y_m$ at the receiver using a hybrid combiner $\mathbf{w}_m$ of size $N_{\mathrm{R}} \times 1$ can be presented as

$$\begin{aligned} y_m &= \sqrt{\rho}\mathbf{w}_m^H \mathbf{H}\mathbf{f}_m s_m + \mathbf{w}_m^H \mathbf{n} \\ &= \sqrt{\rho}\left(\mathbf{f}_m^T \otimes \mathbf{w}_m^H\right) \mathrm{vec}\left(\mathbf{H}\right) + \mathbf{w}_m^H \mathbf{n}. \end{aligned} \tag{6}$$

Note that the mmWave MIMO channel matrix $\mathbf{H}$ can be well approximated by the extended virtual channel model with sufficiently quantized AoAs/AoDs [141]. Therefore, $\mathrm{vec}\left(\mathbf{H}\right)$ can be rewritten as $\mathrm{vec}\left(\mathbf{H}\right) = \mathbf{A}_{\mathrm{D}}\mathbf{h}_{\mathrm{b}}$ where $\mathbf{h}_{\mathrm{b}} = \mathrm{vec}\left(\tilde{\mathbf{H}}_{\mathrm{b}}\right)$ of size $G^2 \times 1$ is a sparse channel vector, and $G$ is number of quantized AoAs/AoDs. The position of each nonzero element in $\mathbf{h}_{\mathrm{b}}$ indicates the AoA and AoD of one channel path, and the value of the nonzero element is the corresponding complex gain of this path. $\mathbf{A}_{\mathrm{D}}$ of size $N_{\mathrm{T}}N_{\mathrm{R}} \times G^2$ is the dictionary matrix of quantified AoAs/AoDs. After receiving $M$ pilots, we have

$$\mathbf{y} = \sqrt{\rho}\begin{bmatrix} \mathbf{f}_1^T \otimes \mathbf{w}_1^H \\ \mathbf{f}_2^T \otimes \mathbf{w}_2^H \\ \vdots \\ \mathbf{f}_M^T \otimes \mathbf{w}_M^H \end{bmatrix} \mathbf{A}_{\mathrm{D}}\mathbf{h}_{\mathrm{b}} + \mathbf{n}_{\mathrm{eff}} = \sqrt{\rho}\boldsymbol{\Psi}\mathbf{A}_{\mathrm{D}}\mathbf{h}_{\mathrm{b}} + \mathbf{n}_{\mathrm{eff}}, \tag{7}$$

where $\mathbf{y} = \left[y_1, y_2, \cdots y_M\right]^T$ and $\mathbf{n}_{\mathrm{eff}}$ is the effective noise vector.

To estimate $\mathbf{h}_{\mathrm{b}}$ from (7), there are two primary approaches proposed in the literature. The first one is to combine the idea of beam-training with sparse signal recovery. For example, in [141], an adaptive channel estimation scheme is proposed. This scheme divides the total channel estimation problem into several subproblems, each of which only considers one channel path. For each channel path, it first starts with a coarse AoA/AoD grid, and determines the AoA and AoD of this path belonging to which angle range by employing OMP algorithm. Then, the AoA/AoD grid around the determined angle range is narrowed and the AoA and AoD of this path are further refined also by the OMP algorithm. In each step, $\mathbf{f}_m$ and $\mathbf{w}_m$ are designed based on the corresponding $\mathbf{A}_{\mathrm{D}}$ (which varies due to different AoA/AoD grid in each step) to make the effective sensing matrix $\boldsymbol{\Psi}\mathbf{A}_{\mathrm{D}}$ have a fixed gain in a specific angle range. This requirement is the same as that of the hierarchical beam-training introduced above. Therefore, a good multi-resolution codebook such as [145], [146] can greatly improve the performance. It has been shown that the adaptive channel estimation scheme enjoys high accuracy with reduced pilot overhead. Some improved schemes following the similar idea can be found in [146], [149]. The second approach is to regard (7) as a classical sparse signal recovery problem [150]–[152]. Then, it designs $\mathbf{f}_m$ and $\mathbf{w}_m$ to make the sensing matrix $\mathbf{\Psi A}_{\rm D}$ enjoy sufficiently low mutual coherence, which is crucial to achieve high signal recovery accuracy in CS theory [153]. In [154], several design methods have been proposed to achieve this goal. Finally, (7) can be solved by lots of efficient CS algorithms, such as OMP and LASSO [153], and the structural sparsity of mmWave MIMO channels (e.g., common support or partial common support) can be also exploited to further reduce the pilot overhead and improve the recovery accuracy [155].

[6]When multiple RF chains are used, the transmitted pilots on different RF chains can be designed to be orthogonal to simplify the channel estimation at the receiver.

### *C. Channel Tracking*

Since the mmWave channel varies over time in mobile networks, the conventional real-time channel estimation should be executed frequently. This has two important consequences [156]: i) In mmWave bands, the user mobility leads to huge Doppler effects and very limited channel coherence time. This means that the mmWave MIMO channels will vary quickly, even if we consider the short symbol duration associated with the wide bandwidth; ii) In hybrid mmWave implementations, there is no enough time to continuously redo the beam-training from scratch. Hence, channel-tracking exploiting the temporal correlation of the time-varying channels is preferable [156]. In this subsection, we describe two main categories of channel-tracking schemes.

The first channel-tracking approach is an improved version of beam-training. The key idea is to select several candidate beam pairs instead of only the optimal beam pair, during each beam-training procedure [142], [157]. For example, when the beam pair with the highest SNR is selected, the beam pairs achieving the second and third highest SNRs are also retained. When the channel varies, only the candidate beam pairs are tested and switched on to keep the SNR above a certain threshold. If all the candidate beam pairs fail, the complete beam-training will be executed again. This idea involves quite low complexity, which makes it easy to implement. Thus, it has been applied in current commercial mmWave communications, such as WLAN (IEEE 802.11ad) [114]. However, this idea is only efficient for single-stream data transmission. For multi-stream data transmission, the beam training itself will incur high pilot overhead, not to mention the search among all candidate beams for all data streams.

An alternative solution to track the mmWave MIMO channel is to utilize the geometric relationship between the transmitter and receiver to track the LoS path of the channel. Specifically, in [156], a priori aided channel-tracking scheme was proposed. By considering a motion model, this scheme first excavates a temporal variation law of the AoA and AoD of the LoS path. After that, by combining the temporal variation law with the sparse structure of mmWave MIMO channels, it utilizes the obtained channels in the previous time slots to predict the prior information, i.e., the support of the channel, in the following time slot without channel estimation. Finally, with the known supports, the time-varying channels can be tracked with a low pilot overhead. The related schemes can be found in [160], [161]. Such schemes can perform well for the LoS path. For the NLoS paths caused by complicated scattering, it is difficult to analyze the geometrical relationship. As a result, a more promising solution in practice is to utilize the idea of the second category to track the LoS path. Then, by eliminating the influence of this path, the NLoS paths can be tracked following the idea of classical Kalman filters [162].

To realize mmWave mobile networks, the efficiency of the beam-tracking schemes must be carefully tested in real environments, to understand which mobility speeds that can be supported and which channel characteristics that can reliably imposed to aid the tracking procedure.

### *D. Hybrid Precoding and Combining*

After the CSI has been acquired, we can design the precoding and combining in the hybrid architecture to achieve the multiplexing gain and array gain offered by MIMO. However, the precoder/combiner design is considerably different from the precoding/combining optimizations in [163] for the conventional fully digital MIMO systems, due to the special hardware constraints in hybrid architecture of mmWave systems. In this subsection, we discuss how to design the hybrid precoder $\mathbf{F} = \mathbf{F}_{\rm A}\mathbf{F}_{\rm D}$ and combiner $\mathbf{W} = \mathbf{W}_{\rm A}\mathbf{W}_{\rm D}$ for N1-N3 architectures discussed in Section III-A with perfect CSI, which helps understand the fundamental limits of the mmWave MIMO with a hybrid architecture. The practically more relevant case of imperfect CSI is largely open and deserves much attention in the upcoming years; in particular, because it is well-known from conventional MIMO communications that precoding/combining schemes that work well under perfect CSI can be widely different from the schemes that work well in practice.

Let us for simplicity consider a non-fading channel $\mathbf{H}$. Specifically, we focus on the optimization problem of hybrid precoding and combining with the aim of maximizing the achievable rate, given by [117]

$$\begin{aligned}\left(\mathbf{F}^{\rm opt}, \mathbf{W}^{\rm opt}\right) &= \arg\max_{\mathbf{F},\mathbf{W}} \log_2 \left|\mathbf{I} + \rho\mathbf{R}_n^{-1}\mathbf{W}^H\mathbf{H}\mathbf{F}\mathbf{F}^H\mathbf{H}^H\mathbf{W}\right|, \\ &\text{s.t. } \mathbf{F}_{\rm A} \in \mathcal{F}, \mathbf{W}_{\rm A} \in \mathcal{W}, \|\mathbf{F}\|_F^2 \leq N_{\rm D}, \end{aligned} \tag{8}$$

where $\mathbf{R}_n$ is the noise/interference covariance matrix, while $\mathcal{F}$ and $\mathcal{W}$ are the sets with all possible analog precoders and combiners satisfying the hardware constraints (which are different for N1-N3 as discussed in Section IV-A), respectively. Generally, obtaining the optimal solution to (8) is a non-trivial task, since $\mathcal{F}$ and $\mathcal{W}$ are much different from conventional fully digital communication. To address (8), several hybrid precoding/combining schemes have been proposed in the literature to achieve feasible solutions.

We first discuss the hybrid precoding and combining schemes for the architecture N1. One effective approach is to decompose the original optimization problem into several subproblems, and each subproblem is approximated as a convex one and then solved by standard convex optimization algorithms. Particularly, a spatially sparse precoding scheme based on the orthogonal pursuit matching (OMP) has been proposed in [117], which can fully exploit the sparsity of mmWave MIMO channel and achieve the near-optimal performance. Using a similar idea, there are some other advanced schemes proposed for N1 [164], [166], [167].

For the subconnected architecture N2, a few hybrid precoding solutions have been developed for maximizing the achievable rate or the spectral efficiency [121], [168]. More recently, the energy efficiency optimization was studied and a solution was proposed in [165]. Since the subconnected architecture adopts a subarray structure, it is natural to solve the complicated hybrid precoding problem by decomposing it into several subproblems and optimizing them in an alternating manner. In particular, a successive interference cancelation (SIC)-based precoding scheme was proposed in [121], which enjoys higher energy efficiency than the spatially sparse precoding scheme. The hybrid precoding problem with energy efficiency as the objective is more complicated. In [165], this optimization problem is solved by jointly exploiting the interference alignment and fractional programming. First, the analog precoder and combiner are optimized via the alternating direction optimization method. Then, the digital precoder and combiner without hardware constraints are obtained based on the effective channel matrix $\mathbf{W}_{\mathrm{A}}^{H}\mathbf{H}\mathbf{F}_{\mathrm{A}}$.

Finally, for the hybrid precoding/combining schemes for N3, the main difficulty lies in the designs of $\mathbf{F}_{\mathrm{A}}$ and $\mathbf{W}_{\mathrm{A}}$, as $\mathbf{F}_{\mathrm{D}}$ and $\mathbf{W}_{\mathrm{D}}$ can be easily obtained based on $\mathbf{W}_{\mathrm{A}}^{H}\mathbf{H}\mathbf{F}_{\mathrm{A}}$. Therefore, it is essential to design an appropriate selecting matrix to select DFT columns (beams) to form $\mathbf{F}_{\mathrm{A}}$ and $\mathbf{W}_{\mathrm{A}}$. Following this idea, a magnitude maximization (MM) beam-selection scheme was proposed in [169], where several beams with large power are selected. Alternatively, an interference-aware beam-selection scheme was proposed in [127]. The key idea is to classify all users into two user groups according to the potential inter-beam interference. For users with low inter-beam interference, it directly selects the beams with large power, while for users with severe inter-beam interference, a low-complexity incremental algorithm was proposed. More beam-selection schemes for N3 can be found in [170], [171].

It worth pointing out that although we have only discussed the single-user multi-stream scenario, the schemes designed for N3 can be extended to multi-user multi-stream scenario. In principle, we can replace the digital SVD precoder by the classical zero-forcing (ZF) precoder to suppress multi-user interference. However, such an extension is not straightforward for the schemes designed for N1 and N2. To tackle this problem, in [172], a two-stage precoding scheme was proposed for architecture N1. In the first stage, $\mathbf{F}_{\mathrm{A}}$ is searched from a predefined codebook to maximize the desired signal power of each user. In the second stage, a digital precoder similar to ZF precoder is designed to cancel multi-user interference. The multi-user hybrid precoding scheme for N2, however, is still an open problem requiring further research efforts.

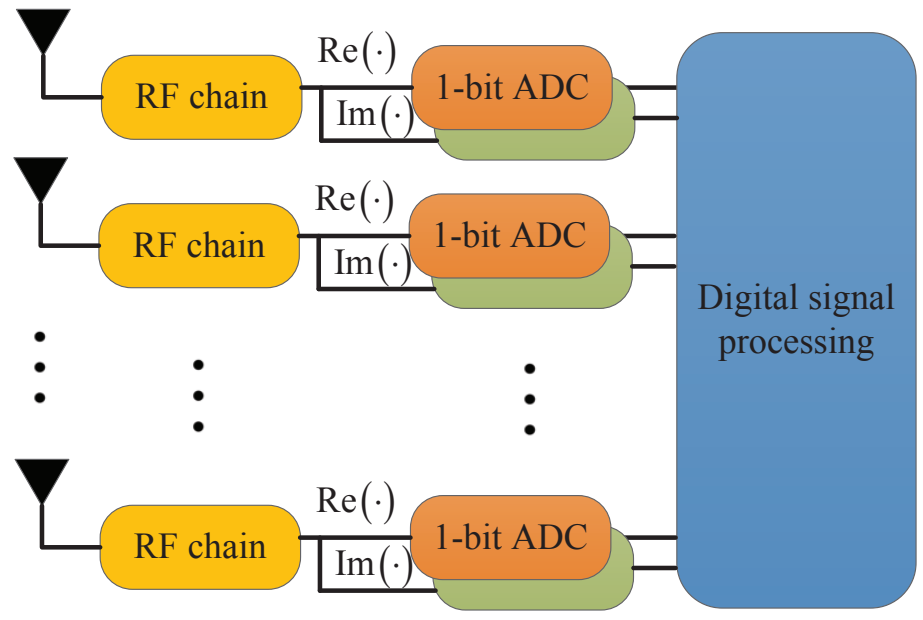

Fig. 9. 1-bit ADC based architecture at the receiver.

### *E. Low-Resolution ADC Based Architecture*

Besides the promising hybrid analog-digital architecture discussed above, some other advanced architectures have also been proposed to reduce the RF complexity of mmWave MIMO systems. The low-resolution (e.g., 1-bit) ADC based architecture[7] is one typical example as shown in Fig. 9 [131]. Different from hybrid analog-digital architecture which aims to reduce the number of RF chains, the key idea of the low-resolution ADC based architecture is to replace the high-resolution (e.g., 15-bit) ADCs by low-resolution ADCs to reduce the energy consumption and hardware cost, while the total number of RF chains or ADCs is still the same as that in the fully digital architecture [131].

The main advantage of low-resolution ADC based architecture is that it can significantly reduce the energy consumption and hardware cost since it only requires a small number of compactor in ADC [132]. Moreover, this architecture can also simplify other circuit modules such as the automatic gain control (AGC) [132]. It has been proved that low-resolution ADC based architecture can achieve the capacity-approaching performance in the low and medium SNR region [133]. However, due to the severe quantization effect of low-resolution ADCs, the capacity of this architecture is usually limited in the high SNR region [133] In addition, the nonlinearity of the quantization also imposes new challenges on the signal processing designs. Take the 1-bit ADC for example, the sampled signal in the digital domain can only be one of the two discrete values instead of one continuous value. To solve this problem, some promising solutions have been proposed. For example, in [134], a channel estimation method using expectation-maximization (EM) algorithm is proposed for low-resolution ADC based architecture to find the maximum a posteriori probability estimate. In [135], a sum-product-algorithm (SPA) based signal detector is designed by utilizing the concept of clustered factor graph. In [136], the authors develop an efficient algorithm for optimal Bayesian data detection in the mmWave OFDM system with low-resolution ADCs. A power allocation (PA) scheme is also proposed to minimize the average symbol error rate in [136].

Nevertheless, besides these works above, there are still some open issues for the low-resolution ADC based architecture [131], e.g., how to design signal processing for broadband mmWave MIMO channels, which require future investigation.

[7]Note that the DACs at the transmitter usually consume less power than ADCs for mmWave MIMO systems. Therefore, employing low-resolution ADCs instead of DACs is more promising to reduce the energy consumption and hardware cost [131].

### *F. Recent Progress*

In the current special issue of IEEE JSAC, the signal processing schemes on mmWave MIMO are investigated in [174]–[189]. Specifically, in [174], a receive antenna selection (RAS)-aid spatial modulation scheme is proposed to reduce the RF complexity, and an iterative algorithm is proposed to search the antenna index with low complexity. In [175], a channel estimation scheme is proposed by utilizing the structured sparsity of mmWave MIMO channel, and a training sequence (TS) is designed by the genetic algorithm. In [176], a low complexity non-iterative interference alignment (IA) scheme for multi-cell mmWave MIMO systems is proposed. In [177], a CANDECOMP/PARAFAC decomposition-based channel estimation scheme is proposed for wideband mmWave MIMO by utilizing the tensor theory. The Cramer-Rao bounds of the estimated channel parameters are also derived to show the advantages of the proposed scheme. In [178], a new antipodal curvedly tapered slot antenna is designed to generate circularly polarized field. It can achieve high gains with low hardware complexity in E-band and W-band. [179] proposes a single-user multi-beam transmission scheme in the beamspace, and the corresponding multi-beam selection, cooperative beam tracking, multi-beam power allocation, and synchronization are investigated. [180] develops a low-complexity channel estimation for hybrid mmWave MIMO systems, and investigates the achievable sum-rate of zero forcing with the estimated channel under the consideration of system imperfection. [181] investigates the hybrid precoding for wideband mmWave MIMO, and proposes a unified heuristic design for two different hybrid precoding architectures, i.e., the fully-connected and the partially-connected architectures, to maximize the overall spectral efficiency. [182] proposes a novel unified hybrid precoding design for fully- and partially-connected hybrid architectures from the view of energy efficiency instead of spectral efficiency. [183] characterizes the gains of pilot precoding and combining in terms of channel estimation quality and achievable data rate. [184] considers the sub-28 GHz communications that do not exhibit enough directivity and selectively, and tackles the sum-rate maximization problem based on the concept of difference of log and trace (DLT) bound. [185] analyzes the achievable rate and energy efficiency of hybrid precoding receivers with low resolution ADC, and shows it robustness to small automatic gain control imperfections. [186] investigates the hybrid precoding design with antenna selection, and decomposes the whole problem into three sub-problems which are solved via an alternating optimization method. [187] proposes a hybrid precoding scheme based on Kronecker decomposition for multi-cell multi-user mmWave MIMO systems. [188] investigates beamforming training for partially-connected hybrid architectures, and proposes two multi-resolution time-delay codebooks. In [188], fundamental limits of beam alignments are studied under different search approaches (exhaustive or hierarchical).

## V. Access, Backhauling, and Coverage

### *A. Multiple-access Technologies*

Multiple-access technologies are necessary for supporting multiple users in mobile networks, and they have been widely investigated in the lower frequency bands. Different multiple access technologies have been utilized in practical systems, including frequency division multiple access (FDMA), time division multiple access (TDMA), code division multiple access (CDMA), and orthogonal frequency division multiple access (OFDMA). These multiple-access technologies are also applicable to mmWave, but in different flavors than in lower frequency bands due to the increased complexity caused by the greatly increased bandwidth, and the different channel characteristics in mmWave bands, e.g., highly directional transmissions. More importantly, as described in Section IV, the shorter wavelengths at mmWave frequencies make MIMO technology suitable for mmWave, since it is possible to use more antennas in the same physical space. One of the key advantages of MIMO is the spatial resolution and beamforming capability that it provides. In this subsection, major non-orthogonal multiple-access technologies that exploit the spatial or power domains will be discussed in the context of mmWave. This includes spatial division multiple access (SDMA), non-orthogonal multiple access (NOMA), and finally random access.

*1) SDMA:* In addition to the orders-of-magnitude larger bandwidths, as compared to conventional systems operating below 6 GHz, a multiplexing gain can be achieved by multiplexing of users in the spatial domain. This technology is known as SDMA was introduced in 1990s [192], and Massive MIMO is the latest branch of this tree [36]. 802.11ac was the first major wireless standard that integrated SDMA [193]. SDMA can ideally increase the sum rate proportionally to the number of multiplexed users, provided that the BS is equipped with at least as many RF transceiver chains as all the users have in total. Due to the highly directional transmissions in mmWave communication systems, users from different directions may be well separated using different spatial beams, which is also known as favorable propagation [191]. The design of transmit precoding/beamforming and receive combining for SDMA is generally a mature topic [194], [195], but the special channel estimation and channel characteristics of hybrid systems require some further work on this topic [196], [197].

One key difference between systems operating in mmWave band and at lower frequencies is the coverage area, which is substantially smaller in mmWave. Hence, there is generally fewer users to multiplex by SDMA in mmWave. Nevertheless, one of the critical challenges of SDMA is how to serve multiple users when the number of users is larger than the number of antennas. It is necessary to group users so that users from different groups may access the BS at the same time, while not causing significant interference to each other. After realizing the user grouping, user scheduling should also be considered to select the users from different groups to access the BS at the same time. Users in the same group can be served orthogonally (or semi-orthogonal) in time, frequency, or code domain. A few effective scheduling algorithms have been proposed in [202]–[210] to enhance throughput.

*2) NOMA:* An alternative to the non-orthogonal spatial multiplexing provided by SDMA is to perform non-orthogonal multiple access (NOMA) in the power domain. This approach is considered as one of the candidates for improving the spectral efficiency and connectivity density in 5G [223]–[227]. At the BS, different signals intended for different users are superimposed on each other after classic channel coding and modulation. Multiple users share the same time-frequency resources, and are then detected at the receivers by successive interference cancellation (SIC). In this way, the spectral efficiency can be enhanced at the cost of an increased receiver complexity compared to conventional orthogonal multiple access (OMA), where the potential interference is treated as noise. NOMA has been considered in mmWave communications in recent literatures. Specifically, the performance of NOMA in mmWave communications was evaluated in [228] and [229], and simulation results have shown that NOMA can achieve better channel capacity than OMA in both uplink and downlink mmWave communication systems. Furthermore, the capacity performance of NOMA-mmWave-massive-MIMO systems was investigated in [230], and simulation results indicated that enormous capacity improvement can be achieved compared to the existing LTE systems. In addition, sum rate and outage probabilities of mmWave-NOMA systems with random beamforming were analyzed in [231], where two users can be served by NOMA in each beam. Furthermore, a transmission scheme that uses NOMA in mmWave beamspace MIMO has been proposed in [232]. By using intra-beam superposition coding and SIC under the framework of NOMA in the proposed beamspace MIMO-NOMA system, more than one user can be simultaneously supported in each beam, which is different from conventional SDMA, where only the signal intended for one user is transmitted in each beam. Consequently, the number of served users can be significantly increased for a given number of beams and RF chains, and the system achievable sum rate can be also improved. Besides, beam division multiple access (BDMA) with per-beam synchronization (PBS) in time and frequency for wideband massive MIMO transmission over mmWave/Terahertz (THz) bands has been proposed in [233], beam scheduling for both UL and DL BDMA and a greedy beam scheduling algorithm has also been developed. Additionally, the design challenges of NOMA in mmWave due to beamforming was investigated in [234]. Note that with the enhanced pathloss in mmWave communications, the interference experienced by the users in NOMA can be significantly reduced [228]. Considering this harmony between the mmWave channel characteristics and the principle of NOMA, the use of NOMA for mmWave communications is a research direction deserves further investigation.

*3) Random access:* This is primarily used for initial access and handover, which is very important in system design. Since random access cannot fully benefit from beamforming due to the lack of information on the best transmit-receive beam pair, the design of the random access channel becomes more challenging in mmWave communication systems [211]–[213]. An overview of random access in mmWave mobile networks has been presented in [211], where the important issues for the design of a random access channel with respect to initial access, handover, uplink-downlink configuration, and scheduling have been discussed in detail.

More recently, there are some recent works emphasize on lower frequencies in ad hoc wireless network scenarios or, more recently, on the 60 GHz IEEE 802.11ad WLAN and WPAN scenarios [214]. While the authors in [215] considered an exhaustive method which sequentially scans the 360 degree angular space, the authors in [216] proposed the use of directional cell discovery procedure where the angular space is scanned in time-varying random directions using synchronization signals periodically transmitted by the base stations. In [217], different scanning and signaling methods were compared, with respect to the initial access design options, in order to assess the respective access delays and system overheads. The analysis shows that low-resolution fully digital architectures have significant benefits when compared with single-stream analog beamforming. In order to reduce the delay due to exhaustive search procedures, the authors in [211], [217], [218] implemented a faster user discovery method which employs a two-stage hierarchical procedure, while the use of context information about user and/or BS positions, provided by a separate control plane, was considered in [214], [219]. In a refinement to the method in [219], the procedure to capture the effects of position inaccuracy and obstacles was implemented in [220]. In addition, the use of booster cells which operate at mm-Waves and under the coverage of a microwave-based anchor cell was proposed by the authors in [221]. In this arrangement, the booster BS gets information about user locations from the anchor BS, enabling it to directly steer the transmit beams towards the user position. Finally, the authors in [222] showed that the performance of analog beamforming degrades in presence of errors in the available context information provided during the initial cell search procedures.

### B. Backhauling

To meet the aggressive 5G KPIs discussed in Section I, an UDN deployment is considered as a promising system architecture to enable Gbps user experience and seamless coverage in mobile networks [235], [236]. In other words, many small-cell BSs are densely deployed as hotspots (e.g., in office buildings, shopping malls, residential apartments) that would greatly offload the macro cells. Hence, the backhaul between the macro-cell BS and the associated small-cell BSs should provide large bandwidth with reliable link transmission. Besides, energy efficiency and deployment cost are also key considerations for operators. It has been demonstrated that backhaul links with 1–10 Gbps is required to effectively support UDN [235]. Conventional optical fiber supports large data rates and reliable link transmission, but its application to UDN as backhaul may not be an economical choice for operators due to the restriction of deployment and installation [237]. Hence, wireless backhaul is more attractive to overcome the geographical constraints. Microwave backhaul in sub-6

GHz and 20 GHz bands have been successfully deployed in current mobile networks. However, the available licensed spectrum in these bands are limited, and insufficient to meet the demand of 5G backhauling, and thus mmWave with wider bandwidth is preferred.

Using mmWave bands for backhauling in UDN is desirable due to the following reasons [237].

- **Large bandwidth**: The large amount of underutilized mmWave including unlicensed V-band (57–67 GHz) and lightly licensed E-band (71–76 GHz and 81–86 GHz) (the specific regulation may vary from country to country) can provide potential GHz transmission bandwidth. For example, more than 1 Gbps backhaul capacity can be supported over a 250 MHz channel in the E-band [238].
- **Reduced interference**: The coverage distance for E-band is up to several km due to rain attenuation, while that for V-band is about 50–700 m due to both the rain and oxygen attenuation. Due to high pathloss, mmWave is suitable for UDN, where improved frequency reuse and reduced inter-cell interference are expected. It should be pointed out that rain attenuation is not a big issue for mmWave used for backhaul in UDN. For example, if we consider very heavy rainfall of 25 mm/h, the rain attenuation is only around 2 dB in the E-band for a backhaul link of 200 m [238].
- **In-band backhauling**: The backhaul and user access links are conventionally carried out in different frequency bands. Multiplexing backhaul and access on the same mmWave frequency band, also named in-band backhaul [235], has obvious cost benefits from the hardware and frequency reuse perspective. However, as in any in-band backhauling scenario, the interference between the backhaul and the user access links must be controlled; for example, by beamforming and signal processing.

### *C. Coverage and Connectivity*

Since mmWave signals in general have high penetration loss, they are very sensitive to the blockage by walls and other objects. In [242], a mathematical framework for modeling the random blockages that occur when users are moving around was developed by leveraging concepts from random shape theory. In this model, both BSs and users are assumed to form a homogeneous Poisson point process (PPP) on the plane. Random buildings are modeled as rectangles with random sizes and orientations whose centers form a PPP on the plane. Let $K$ denote the total number of blockages crossing a link. As shown in [242], $K$ follows a Poisson distribution with the mean of $\beta R + p$, where $\beta = 2\lambda(E[W] + E[L])/\pi$, $p = \lambda E[L]E[W]$, $E[L]$ and $E[W]$ represent the average length and width of a random blockage building, respectively. Then, the probability that link of length $R$ is free from blockage has an exponential distribution, given by

$$P(K = 0) = e^{-(\beta R + p)}. \tag{9}$$

It should be noted that in practical mmWave systems the parameters $\beta$ and $p$ in (9) need to be obtained through experimental fitting.

A visible LoS region of location $X$ is defined as the set of locations which can be connected to $X$ with a LoS link. If we consider the case where the blockages are impenetrable, the average size of a visible region in a cellular network can be shown to be $2\pi e^{-p}/\beta^2$ [242] and the average number of BSs which has a LoS link with a user is $2\pi\mu e^{-p}/\beta^2$, where $\mu$ is the density of the BS PPP. Therefore, in order to achieve acceptable coverage, we can either increase the density of the BS deployment or reduce the physical length of the communication links between two nodes in the network by deploying more intermediate relays. The paper [173] systematically studies the blockage problem of mmWave networks. The impact of reflection and relaying are also investigated. It is shown that if the LoS signals are blocked, the reflection signals may be useful in regions, while for some regions relaying is preferred. Reference [173] also studies the optimal routing problem in multiple relay mmWave networks. Recently, an optimal opportunistic strategy for access point deployment is proposed in [247] to well balance the overhead and capacity under random blockage.

[243] takes a stochastic geometry approach to the connectivity of mmWave networks with multi-hop relaying. It is shown that multi-hop relaying can greatly improve the connectivity compared to the single-hop mmWave transmission. The results also show that to obtain near-optimal connectivity the relaying route window should be about the size of the obstacle buildings. The coverage probability is an important performance metric in a mmWave cellular network. It is defined as the probability that the destination is able to receive a signal with a certain threshold SNR $T$:

$$P_c(T) = \Pr(\text{SNR} > T). \tag{10}$$

To further improve the coverage, [244] studies the possibility of BS cooperation in the downlink of mmWave networks in a stochastic geometry framework. It is shown that cooperation among randomly located BSs can effectively increase the coverage probability.

By using tools from stochastic geometry, a general and tractable framework for coverage analysis with arbitrary distributions for interference power and arbitrary antenna patterns was developed and applied to mmWave ad hoc networks exploiting directional antenna arrays in [245]. It is shown that the coverage probabilities of mmWave ad hoc networks increase as a non-decreasing concave function with the antenna array size. Numerical results also show that largescale antenna arrays are required for satisfactory coverage in mm-wave networks. To further enhance the connectivity and session continuity, multi-connectivity strategies were developed to leverage multiple simultaneous small cell connections in ultradense urban deployments of mmWave networks in [246]. The benefits of multi-connectivity strategies were investigated by taking into account: (i) the intricacies of mmWave radio propagation in realistic urban environments; (ii) the dynamic mmWave link blockage due to human mobility; and (iii) the multi-connectivity network behavior to preserve session continuity. Results show that even simpler multi-connectivity schemes bring notable improvements to session-level mmWave operation in realistic environments.

To achieve a robust coverage and connectivity of mmWave networks, a heterogeneous mm-wave network architecture consisting of small-cell BSs operating at mmWave bands and macro-cell BSs, operating at microwave frequencies, should be considered in practice. The macro-cells will be used as a signaling network for cell discovery and signaling transmission to guarantee full coverage and provide reliable control channels. The small-cell BSs form a data subnetwork that provides high data rates using mmWave bands for the users that they cover.

## VI. Standardization and Deployment

Despite of the huge potential and great number of benefits in using mmWave in mobile networks, there are still a lot of skepticisms, particularly from investors, as for whether the technology is suitable for cellular coverage and mobility scenarios. Due to the great potential of mmWave communications as described in previous sections, 3GPP is working towards standardizing mmWave for in the 5G New Radio (NR) interface. More specifically, 3GPP is working on Release 14 (will be finalized around June 2017), which will include the channel modeling for radio above 6GHz. From the second half of 2017, 3GPP will work on release 15, which will deliver the first set of 5G standards. Commercial use of the unlicensed mmWave band is not a new territory, but mmWave bands have been used since the 1980s. However, the existing standards and applications have only been for static ultra-high-definition multimedia data transfer. Specifically, IEEE 802.11ad/aj for wireless local area network (WLAN), IEEE 802.15.3c and ECMA-387 for wireless personal area network (WPAN), WirelessHD for video area networking (VAN) are standards at unlicensed 60GHz and 45GHz for providing short-range point-to-point (P2P) communications.

In this section, we will introduce the key features of the aforementioned existing and in-development mmWave standards, followed by an overview of the plan for commercial cellular deployment.

### A. 3GPP's New Radio at mmWave Band

*1)* ***Vision and use cases****:* 5G is envisioned to provide orders-of-magnitude improvements in the peak data rate, network capacity, latency, availability, and reliability over legacy networks. Meanwhile, the deployment should cooperate seamlessly with legacy networks and provide fundamental shifts in cost and energy efficiency [120]. As discussed in Section I, 3GPP has defined three use cases for 5G NR since 2016; namely, eMBB, mMTC and URLLC. EMBB is targeted for mobile broadband services that require extraordinary data rates; mMTC is the basis for connectivity in Internet of Things (IoT); and URLLC is needed for applications which have stringent latency and reliability requirements. 3GPP has also identified its interest in the frequency bands above 6 GHz and up to 100 GHz for NR. The channel model for above 6 GHz is defined in 3GPP Specification 38.900/38.901 [21], [98]. Network operators around the world are exploring the possibility of using parts of this spectrum for licensed mobile communications to meet diversified services from their customers. mmWave is naturally envisioned to provide eMBB services where high-speed data transmission is needed. This is particularly important in small cells and dense urban scenarios. Other applications are backhaul for last-mile fiber replacement, wireless fronthauls, etc. However, the research has not been very clear on how mmWave affects the latency and reliability aspects.

*2)* ***mmWave and massive MIMO****:* The 5G vision is pushing for a fundamental change in the mobile networks, starting with the lowest physical layer (PHY). The two core PHY technologies, that will set 5G NR apart from previous radio access technologies (RATs), are mmWave and massive MIMO. Both put forth different paradigms that make a break with many current understandings in the wireless propagation, signal processing, device manufacturing and eventual network design [113]. Moreover, there is a clear trend that mmWave and massive MIMO will work in cooperation to ensure a successful 5G operation. The great SDMA capability of massive MIMO is ideal in the frequency bands below 6 GHz, where a large area is covered and many users with NLoS channels need to be served simultaneously. The high capacity of mmWave communications is ideal for hotspots that need not guarantee coverage or support mobility, but can provide great service to the LoS users that are static or moving at pedestrian speed. In 3GPP, the maximum number of the RF chains for the NR BS and UE is determined as 32 and 8 respectively. The maximum number of antenna elements can go up to 1024 at the BS and 64 at the UE (for 70 GHz), respectively.

*3)* ***Hybrid beamforming Architecture****:* At the PHY layer, the hybrid analog-digital beamforming technique was agreed to be used in 5G systems at the 3GPP RAN1 meeting in 2016 [125]. As discussed in Section IV, the hybrid architecture may be used by both the BSs and UEs. It is worth noting that the hybrid beamforming would not only be used to boost the data rate for data transmission, but also be used at the control channel to enhance cell coverage. Thereafter, it is deemed necessary to have Physical Uplink Control Channel (PUCCH) to include CSI that is related to analog beamforming information. As for the Physical Downlink Control Channel (PDCCH), it is desirable to have a common design for high and low frequency bands. Both uniform antenna arrays (i.e., antenna elements with the same polarization from multiple panels are uniformly distributed in horizontal and vertical dimensions, respectively) and non-uniform antenna arrays [250] (i.e., antenna elements with same polarization from multiple panels are not uniformly distributed in horizontal or vertical dimension) should be supported in 5G systems. This means that a flexible design of the hybrid beamforming is needed, which is not limited to calibrated arrays of a particular geometry. Spatial beams can be generated from separate antenna panels (or different sections of the antenna array) to serve different users. Meanwhile, it is possible to allocate different transmission tasks (i.e., data channel and control channel) to each subarray. Nevertheless, fully digital beamforming is not disregarded by industry, and low-resolution beamforming could work at the lower end of the mmWave spectrum with manageable complexity.

*4)* ***PHY layer design****:* Some common understandings on the way forward for the standardization of the PHY layer for mmWave communications have been agreed upon. More detailed discussion and specifications are still open and are expected to be finalized in the 3GPP Release 15 in 2018. Research should focus on, but not be limited to, the following areas: i) Beam-management, which includes the beam-sweeping procedure, beam-selection based on CSI feedback or beam-reciprocity assumptions, beam-tracking and recovery for mobility support, etc.; ii) the corresponding Reference Signal (RS) design for beam-management; and iii) Control channel design.

TABLE III
SUMMARY OF MMWAVE PERFORMANCE MEASUREMENT CAMPAIGNS

| | Huawei | QUALCOMM | Ericsson | Samsung | Nokia |
|---|---|---|---|---|---|
| Freq band | 28 GHz [252]<br>73 GHz [253] | 28 GHz [254] | 28 GHz [257] | 28 GHz [256] | 28 GHz<br>73 GHz [255] |
| Architecture | Digital (@28 G)<br>Analog (@73 G) | Hybrid | Hybrid | Hybrid | Hybrid(@28GHz)<br>RF(@73GHz) |
| Peak Throughput (Gbps) | 35(@73 G) | 10 | 14 | 7.5 | 11 |

Beam-sweeping is a main method for the analog part of the hybrid beamforming, and its procedure design has a big impact on the system implementation. To name a few, how to design an efficient and reliable sweeping procedure during the initial access stage, how to perform beam-tacking for a UE as the propagation channel changes (due to fast fading or mobility), and how to maintain a connection (i.e., beam-recovery) in the case of link failure and/or blockage. A well designed beam-selection mechanism can ensure good signal strength to reap the benefits of data transmission at the mmWave frequencies with a wide bandwidth. The accuracy of the selection is dependent on the analog beam-feedback, the digital precoding matrix feedback and the Channel Quality Index (CQI) feedback. The common understanding is that the UE will report measurement results on different BS transmit beams to aid beam-selection at the BS.

### *B. Prototypes and Deployment plan*

There has been tremendous efforts from various organizations (vendors, operators, universities, etc.) in building hardware platforms for mmWave channel measurement (as described in Section II) and for proof-of-concept prototyping. This is the necessary step to test for different frequency bands, different use cases, and their respective KPIs before the mass deployment. In the past year of 2016, there have been many feasibility tests and functionality tests for potential key techniques proposed for mmWave. A large quantity of measurement data has been accumulated, which is vital for clarification on the system architecture and the PHY design, as well as the subsequent standardization and commercialization for mmWave communication. Table III lists some known operational prototypes and their key system parameters.

Current measurements are largely focused on four areas of interest.

1) Basic throughput performance: the maximum number of reported spatial streams is two (not counting dual-polarization of antenna elements) and the maximum spectral efficiency is less than 20 bps/Hz.
2) NLoS transmission: metal and concrete walls pose high penetration loss; heavy foliage poses severe shadowing loss; reasonable reflection loss but high diffraction loss.
3) Coverage: indoor $\leq$ 100 m; urban outdoor: $\leq$ 350 m; outdoor-to-indoor: $\leq$ 20 m.
4) Beam-tracking: It can be done for a single user at pedestrian (15 km/h) or low speed (40 km/h); multi-user tracking (spatially) is lacking.

The next phase of testing will focus on the feasibility in more difficult use cases for a single channel; for instance, outdoor BS serving indoor UEs, maximum outdoor coverage test, high-mobility/vehicular scenarios, beam-tracking for multiple UEs, dense urban services, multi-cell networking and core network support. This will require the consumer equipments to be developed and ready for such tests, so that end-to-end performance evaluation can be possible. It is most likely that mass deployment of mmWave technology for mobile networks will occur after 2020, since the progress still largely hinges on the standardization process and the resolving of the aforementioned challenges.

## VII. CONCLUSIONS

Despite of high potentials of providing multiple Gbps rates, many technical challenges have to be solved for mmWave communications to become a mainstream technology in mobile networks. In recent years, large efforts have made to tackle the various challenges and many excellent results have been reported. This tutorial paper has summarized the recent technical progresses in mmWave communications for mobile networks, including channel measure/modeling, MIMO design, multiple access, performance analysis, standardization, and deployment. Many directions for future research work have also been identified. From our point of view, finding effective solutions for applying mmWave technology in high mobility environments, enabling enhanced transmission distance, combating hardware impairments, and achieving high energy-efficiency are very important and interesting challenges to tackle. From a broader perspective, the RF implementation of mmWave technology is very important and a long-term goal should be to obtain a cost-efficient fully digital implementation.